\documentclass{aa}  
\usepackage{graphicx}
\usepackage{txfonts}
\newcommand{\mincir}{\raise -2.truept\hbox{\rlap{\hbox{$\sim$}}\raise5.truept
\hbox{$<$}\ }}
\newcommand{\magcir}{\raise -2.truept\hbox{\rlap{\hbox{$\sim$}}\raise5.truept
\hbox{$>$}\ }}
\newcommand{\siml}{\raise -2.truept\hbox{\rlap{\hbox{$\sim$}}\raise5.truept
\hbox{$<$}\ }}
\newcommand{\simg}{\raise -2.truept\hbox{\rlap{\hbox{$\sim$}}\raise5.truept
\hbox{$>$}\ }}
\newcommand{\be}{\begin{equation}}
\newcommand{\ee}{\end{equation}}
\newcommand{\ba}{\begin{eqnarray}}
\newcommand{\ea}{\end{eqnarray}}
\newcommand {\kpc} {$h_{70}^{-1}$ kpc $\;$}

\newcommand {\h} {$h_{70}^{-1}$ Mpc$\;$}
\newcommand {\hh} {$h_{70}^{-1}$ Mpc}
\newcommand {\hhh} {\;h_{70}^{-1} \mathrm{Mpc}}
\newcommand {\ks} {km~s$^{-1} \;$}
\newcommand {\kss} {km~s$^{-1}$}

\newcommand {\mqui} {$\times 10^{15}\;h_{70}^{-1}\;M_{\odot} \;$}
\newcommand {\mquii} {$\times 10^{15}\;h_{70}^{-1}\;M_{\odot}$}

\newcommand {\mll} {$h_{70}\;M_{\odot}/L_{\odot}$}

\newcommand{\degree}{\ensuremath{\mathrm{^\circ}}}
\newcommand{\arcm}{\ensuremath{\mathrm{^\prime}\;}}
\newcommand{\arcs}{\ensuremath{\arcmm\hskip -0.1em\arcmm \;}}
\newcommand{\arcmm}{\ensuremath{\mathrm{^\prime}}}
\newcommand{\arcss}{\ensuremath{\arcmm\hskip -0.1em\arcmm}}

\newcommand{\dotsec}{\,\rlap{\hbox{$\mathrm{^s}$}}{\hbox{$.$}}\,}

\begin{document}

\title{Abell 1758N from an optical point of view: new insights on a merging cluster with diffuse radio emission} 

   \author{
W. Boschin\inst{1}
          \and
M. Girardi\inst{2,3}
          \and
R. Barrena\inst{4,5}
          \and
M. Nonino\inst{3}
}

   \offprints{W. Boschin, \email{boschin@tng.iac.es}}

   \institute{ 
Fundaci\'on Galileo Galilei - INAF (Telescopio Nazionale Galileo), Rambla Jos\'e Ana Fern\'andez Perez 7, E-38712
     Bre\~na Baja (La Palma), Canary Islands, Spain\\ 
\and Dipartimento di Fisica dell'Universit\`a degli Studi
     di Trieste - Sezione di Astronomia, via Tiepolo 11, I-34143
     Trieste, Italy\\ 
\and INAF - Osservatorio Astronomico di Trieste,
     via Tiepolo 11, I-34143 Trieste, Italy\\ 
\and Instituto de
     Astrof\'{\i}sica de Canarias, C/V\'{\i}a L\'actea s/n, E-38205 La
     Laguna (Tenerife), Canary Islands, Spain\\ 
\and Departamento de
     Astrof\'{\i}sica, Universidad de La Laguna, Av. del
     Astrof\'{\i}sico Francisco S\'anchez s/n, E-38205 La Laguna
     (Tenerife), Canary Islands, Spain
}

\date{Received  / Accepted }

\abstract{The mechanisms producing the diffuse radio emission in
  galaxy clusters, and in particular their connection with cluster
  mergers, are still debated.}{We seek to explore the internal
  dynamics of the cluster Abell 1758N, which has been shown to host a
  radio halo and two relics, and is known to be a merging bimodal
  cluster.}{Our analysis is mainly based on new redshift data for 137
  galaxies acquired at the Telescopio Nazionale Galileo, only four of
  which have redshifts previously listed in the literature. We also
  used photometric data from the Sloan Digital Sky Survey and from the
  Canada--France--Hawaii Telescope archive. We combined galaxy
  velocities and positions to select 92 cluster galaxies and analyzed
  the internal cluster dynamics.}{We estimate a cluster redshift of
  $\left<z\right>=0.2782$ and quite a high line--of--sight (LOS)
  velocity dispersion $\sigma_{\rm V}\sim 1300$ \kss. Our 2D analysis
  confirms the presence of a bimodal structure along the NW--SE
  direction.  We add several pieces of information to the previous
  merging scenario: the two subclusters (here A1758N(NW) and
  A1758N(SE)) cannot be separated in the velocity analyses and we
  deduce a small LOS velocity difference ($\Delta V_{\rm rf,LOS}\siml
  300$ \ks in the cluster rest--frame). The velocity information
  successfully shows that A1758N is surrounded by two small groups and
  active galaxies infalling onto, or escaping from, the
  cluster. Removing the two groups, we estimate $\sigma_{\rm V,NW}\sim
  1000$ \ks and $\sigma_{\rm V,SE}\sim 800$ \ks for A1758N(NW) and
  A1758N(SE), respectively. We find that Abell 1758N is a very massive
  cluster with a range of $M=2-3$ \mquii, depending on the adopted
  model.}{As expected for clusters that host powerful, extended,
  diffuse radio emissions, Abell 1758N is a major cluster merger just
  forming a massive system.}

  \keywords{Galaxies: clusters: individual: Abell 1758 --
             Galaxies: clusters: general -- Galaxies: kinematics and
             dynamics}

  \titlerunning{Abell 1758N: new optical insights on a merging cluster with diffuse radio emission}
   \maketitle
%
%________________________________________________________________

\section{Introduction}
\label{intro}

Merging processes constitute an essential ingredient of the evolution
of galaxy clusters (Feretti et al. \cite{fer02b}). An interesting
aspect of these phenomena is the possible connection between cluster
mergers and extended, diffuse radio sources: halos and relics. The
synchrotron radio emission of these sources demonstrates the existence
of large-scale cluster magnetic fields and of widespread relativistic
particles. Cluster mergers have been proposed to provide the large
amount of energy necessary for electron reacceleration to relativistic
energies and for magnetic field amplification (Tribble \cite{tri93};
Feretti \cite{fer99}; Feretti \cite{fer02a}; Sarazin
\cite{sar02}). Radio relics (``radio gischts'' as they were called by
Kempner et al. \cite{kem04}), which are polarized and elongated radio
sources located in the cluster peripheral regions, seem to be directly
associated with merger shocks (e.g., Ensslin et al. \cite{ens98};
Roettiger et al. \cite{roe99}; Ensslin \& Gopal-Krishna \cite{ens01};
Hoeft et al. \cite{hoe04}).  Radio halos are more likely associated
with the turbulence following a cluster merger, although the precise
radio formation scenario remains unclear (re-acceleration vs. hadronic
models, e.g., Brunetti et al. \cite{bru09}; Ensslin et
al. \cite{ens11}).  Recent semi-analytical calculations in the
framework of the turbulent re-acceleration scenario have allowed the
community to derive the expectations for the statistical properties of
giant radio halos, in agreement with present observations that halos
are found in very massive clusters (Cassano \& Brunetti \cite{cas05};
Cassano et al. \cite{cas06}).  Alternative models have been presented
in the framework of hadronic models, where the time-dependence of the
magnetic fields and of the cosmic ray distributions is taken into
account to explain the observational properties of both halos and
(most) relics (Keshet \& Loeb \cite{kes10}). In these models the
intracluster medium (ICM hereafter) magnetization is triggered by a
merger event, in part but probably not exclusively in the wake of
merger shocks.  Unfortunately, one has been able to study these
phenomena only recently on the basis of a sufficient statistics, i.e.
a few dozen clusters hosting diffuse radio sources up to $z\sim 0.5$
(e.g., Giovannini et al. \cite{gio99}; see also Giovannini \& Feretti
\cite{gio02}; Feretti \cite{fer05a}; Venturi et al. \cite{ven08};
Bonafede et al. \cite{bon09}; Giovannini et al. \cite{gio09}).  It is
expected that new radio telescopes will highly increase the statistics
of diffuse sources (e.g., LOFAR, Cassano et al. \cite{cas10a}).

From the observational point of view, there is growing evidence of the
connection between diffuse radio emission and cluster mergers, since
up to now diffuse radio sources have been detected only in merging
systems (see Cassano et al. \cite{cas10b}). In most cases the cluster
dynamical state has been derived from X-ray observations (Schuecker
et al. \cite{sch01}; Buote \cite{buo02}; Cassano et al.
\cite{cas10b}).  Optical data are a powerful way to investigate the
presence and the dynamics of cluster mergers, too (e.g., Girardi \&
Biviano \cite{gir02}). The spatial and kinematical analysis of member
galaxies allow us to reveal and measure the amount of substructure,
and to detect and analyze possible pre-merging clumps or merger
remnants.  This optical information is indeed complementary to the
X-ray information because galaxies and the ICM react on different
timescales during a merger (see, e.g., the numerical simulations by
Roettiger et al. \cite{roe97}).  In this context, we are conducting an
intensive observational and data analysis program to study the
internal dynamics of clusters with diffuse radio emission by using
member galaxies (DARC -- Dynamical Analysis of Radio Clusters --
project, see Girardi et al. \cite{gir07}\footnote{see also the web
  site of the DARC project
  http://adlibitum.oat.ts.astro.it/girardi/darc.}).

During our observational program, we have conducted an intensive study
of the cluster Abell 1758 (hereafter A1758).  A1758 is a very rich
Abell cluster having Abell richness class $=3$ (Abell et
al. \cite{abe89}).  Optically, the cluster is classified as
Bautz-Morgan class III (Abell et al. \cite{abe89}) and a Rood-Sastry
irregular cluster ``F'', i.e. it has a flattened configuration
(Struble \& Rood \cite{str87}).  While A1758 was classified as a
single cluster by Abell, ROSAT images show that there are two distinct
clusters (\object{Abell 1758N} and Abell 1758S, hereafter A1758N and
A1758S), both highly disturbed systems, separated by approximately 2
\h along the N-S direction (Rizza et al. \cite{riz98}) and both very
X-ray luminous: $L_\mathrm{X}$(0.1--2.4 keV)=$11.68 \times 10^{44}
\ h_{50}^{-2}$ erg\ s$^{-1}$ and $L_\mathrm{X}$(0.1--2.4 keV)=7.25
$\times 10^{44} \ h_{50}^{-2}$ erg\ s$^{-1}$ (Ebeling et
al. \cite{ebe98}, in their cosmology) for A1758N and A1758S,
respectively.

The reference X-ray study for A1758 is that of David \& Kempner
(\cite{dav04}), based on both {\em Chandra} and {\em XMM-Newton}
data.  According to their analysis, A1758N and A1758S do not reveal
any sign of interaction between the two clusters (see also Durret et
al. \cite{dur11}) and their LOS recessional velocity, as measured from
the X-ray spectra, is within 2100 \kss, suggesting that they likely
form a gravitationally bound system. Moreover, David \& Kempner
(\cite{dav04}) found that A1758N is in the late stages of a large
impact parameter merger between two hot ($kT_{\rm X}\sim$7 keV)
subclusters (NW and SE subclumps), with the two remnant cores
separated in projection by 800 \kpc and surrounded by hotter gas
($kT_{\rm X}\sim$9--12 keV) that was probably shock-heated during the
early stages of the merger. In particular, the {\em Chandra} image
suggests that the X-ray NW subclump is currently moving toward the
N, while the X-ray SE subclump is moving toward the SE.

The optical luminosity distribution of the cluster red-sequence
galaxies reveals two luminous NW and SE subclumps, too (see Okabe \&
Umetsu \cite{oka08}). However, while the peak position of the X-ray
NW subclump coincides with that of the luminous brightest cluster
member lying at NW, the X-ray SE subclump is offset by about 290 \kpc
to the NW of the optical SE structure. The bimodal feature of A1758N
in the weak-lensing mass maps reported by Dahle et al. (\cite{dah02})
and Okabe \& Umetsu (\cite{oka08}) shows that the mass and light are
similarly distributed in A1758N. Both studies found arc candidates
around the two mass peaks. The SE structure in the mass/galaxy map is
located in front of the X-ray SE subclump and moves toward the
south-east, while the NW structure shows no significant offset among
the galaxy, ICM, and mass distributions (Okabe \& Umetsu
\cite{oka08}; Ragozzine et al. \cite{rag12}).

More recently, Durret et al. (\cite{dur11}) confirmed the complex
structure of A1758N and the high core X-ray temperature (6-7 keV) with
{\em XMM-Newton} data. They also found two elongated regions of high
metallicity in A1758N, likely left by ram-pressure stripping during
the merger of two subclusters. Moreover, using Canada-France-Hawaii
Telescope (CFHT) images, Durret et al. (\cite{dur11}) computed A1758N
and A1758S luminosity functions and discussed differences from a
Schechter function in the view of the cluster internal dynamics.

\begin{figure*}
\centering 
\includegraphics[width=18cm]{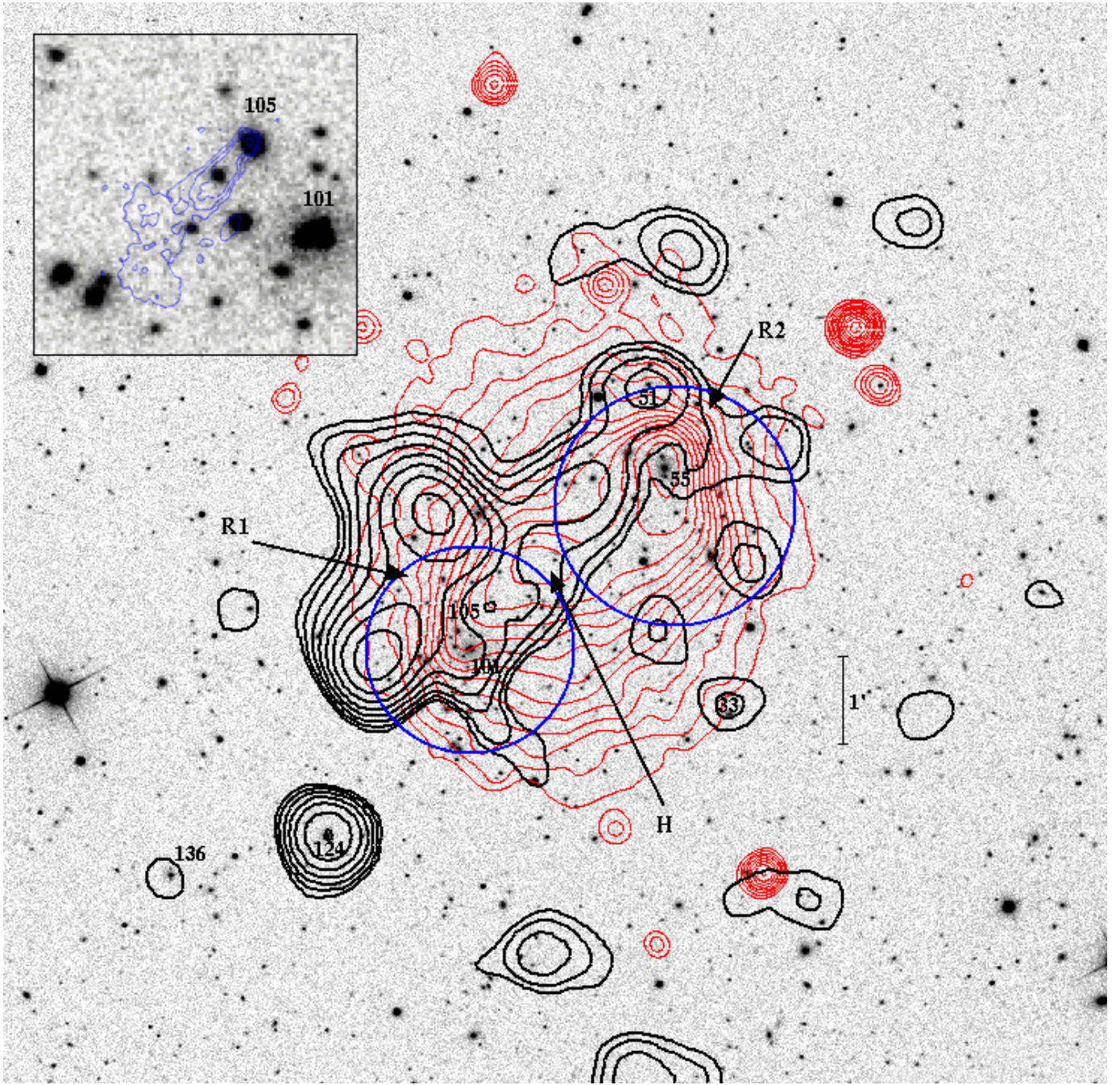}
\caption{Multiwavelength picture of the cluster A1758N (north at the
  top and east to the left). Optical density peaks detected from our
  analysis of the galaxy distribution (big blue circles; see
  Sect.~3.4) are superimposed on the SDSS $r^{\prime}$-band image of
  the cluster region. Labels indicate the IDs of galaxies cited in the
  text. Thin red contour levels show the ICM distribution we derived
  from the X-ray Chandra archival image ID~2213 (photons in the energy
  range 0.3-7 keV; see David \& Kempner \cite{dav04} for an extensive
  analysis of these X-ray data). From Giovannini et al. (\cite{gio09})
  we also reproduce the radio contour levels of their VLA radio image
  at 1.4 GHz (thick black contours, HPBW=45\arcss$\times$45\arcss) and
  highlight with arrows the positions of the radio halo (H, in the
  center) and the two peripheral radio relics (R1 and R2). The insert
  on the top left is a zoom on the region of the NAT radio galaxy
  1330+507 (ID~105), with the radio contours taken from VLA
  high-resolution radio images presented by O'Dea \& Owen
  (\cite{ode85}).}
\label{figimage}
\end{figure*}

As for the diffuse radio emission, the first evidence of a diffuse
radio source in A1758N came from Kempner \& Sarazin (\cite{kem01}).
Giovannini et al. (\cite{gio06}) first reported the clear detection of
a radio halo in the center of the cluster. More recently, Giovannini
et al. (\cite{gio09}) identified a central diffuse radio emission
(halo), 0.63 \h in size, and two brighter peripheral structures on the
opposite sides with respect to the cluster center that resemble relic
radio sources (see their Fig.~7 and our Fig.~\ref{figimage} for a
multiwavelength view of the cluster). A1758N is one of the very few
clusters with a central halo and two peripheral relics similar to
RXCJ1314.4-2515 (see Feretti et al. \cite{fer05b}).

{\em Spitzer/}MIPS 24 $\mu$m observations of A1758 show a very active
star formation in the A1758N galaxies, suggesting that dust-obscured
activity in clusters is triggered by the details of cluster-cluster
mergers, too (Haines et al. \cite{hai09}).

To date, only a small amount of redshift data have been available for
A1758, they indicate a value of $z\sim0.28$. No internal dynamical
analysis based on member galaxies has been published.

Our new spectroscopic data for A1758N come from the Telescopio
Nazionale Galileo (TNG). Our present analysis is also based on public
optical photometric data from the Sloan Digital Sky Survey (SDSS) and
the CFHT archive.

This paper is organized as follows. We present our new optical data
and the cluster catalog in Sect.~2. We present our results about the
cluster structure in Sect.~3.  We briefly discuss our results and
present our conclusions in Sect.~4.

Unless otherwise stated, we indicate errors at the 68\% confidence
level (hereafter c.l.). Throughout this paper, we use $H_0=70$ km
s$^{-1}$ Mpc$^{-1}$ in a flat cosmology with $\Omega_0=0.3$ and
$\Omega_{\Lambda}=0.7$. In the adopted cosmology, 1\arcm corresponds
to $\sim 254$ \kpc at the cluster redshift.

\begin{figure*}
\centering 
\includegraphics[width=18cm]{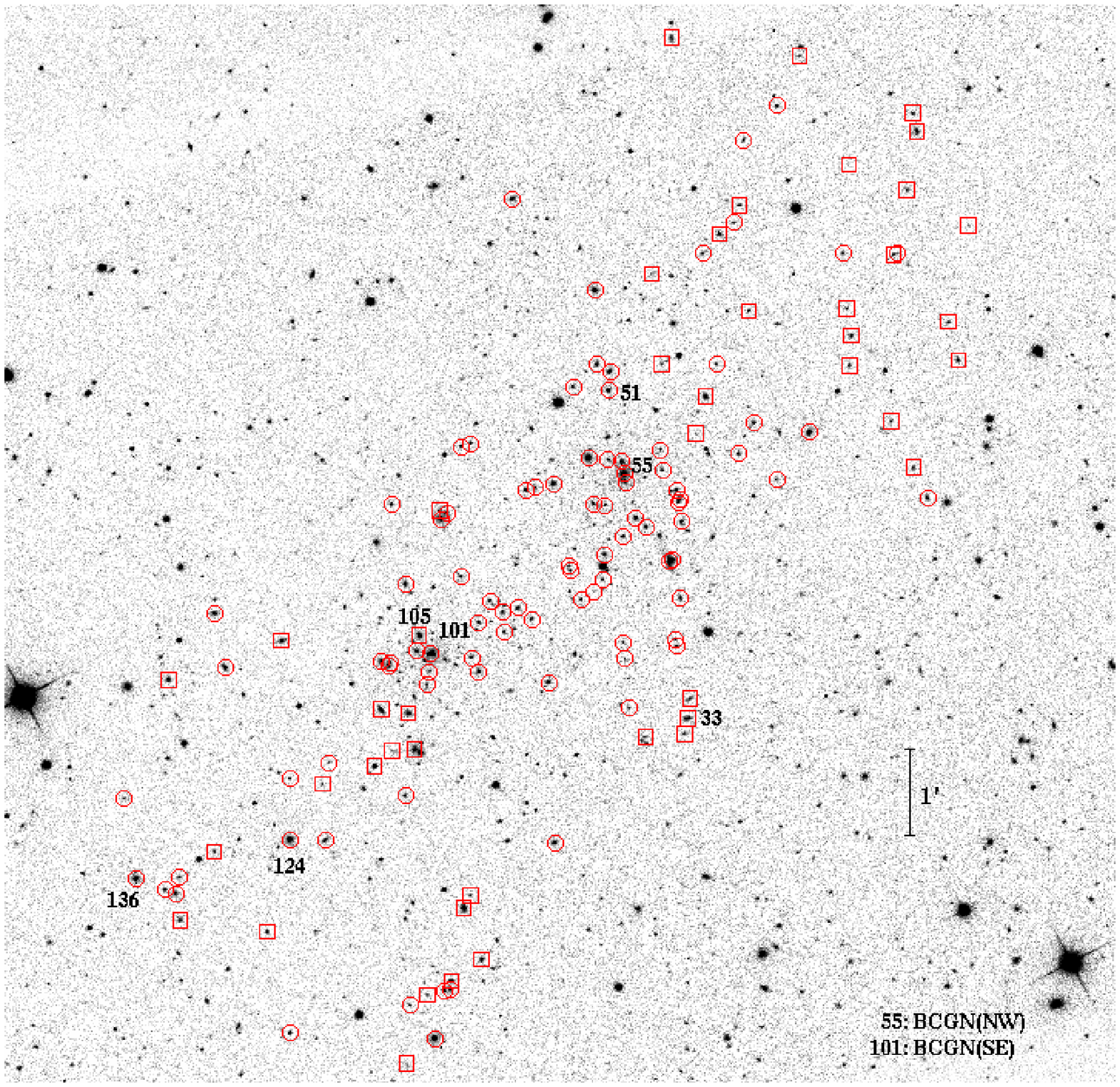}
\caption{ SDSS $r^\prime$-band image of A1758N (north at the top and
  east to the left) showing galaxies with measured spectroscopic
  redshifts. Circles and boxes indicate cluster members and
  nonmember galaxies, respectively (see
  Table~\ref{catalogA1758}). Labels indicate the IDs of galaxies cited
  in the text.}
\label{figottico}
\end{figure*}

\section{Redshift data and galaxy catalog}
\label{data}

Multi-object spectroscopic observations of A1758N were carried out at
the TNG, a 4m-class telescope, in May 2008 and May 2009. We used
DOLORES/MOS with the LR-B Grism 1, yielding a dispersion of 187
\AA/mm. The detector is a $2048\times2048$ pixels E2V CCD, with a
pixel size of 13.5 $\mu$m. In total, we observed four MOS masks (one
in 2008 and three in 2009) for a total of 146 slits. The targets were
chosen within the pre-imaged area (a rectangle of
$\sim$8.6\arcmm$\times$13.7\arcm centered on the cluster and with the
major side clockwise rotated by 40\degree) after a rough color
pre-selection. We acquired three exposures of 1800 s for each
mask. Wavelength calibration was performed using helium+argon and
helium+mercury+neon lamps. Reduction of spectroscopic data was carried
out using the IRAF\footnote{IRAF is distributed by the National
  Optical Astronomy Observatories, which are operated by the
  Association of Universities for Research in Astronomy, Inc., under
  cooperative agreement with the National Science Foundation.}
package. Radial velocities were determined using the cross-correlation
technique (Tonry \& Davis \cite{ton79}) that is implemented in the
RVSAO package (developed at the Smithsonian Astrophysical Observatory
Telescope Data Center). Each spectrum was correlated against six
templates for a variety of galaxy spectral types: E, S0, Sa, Sb, Sc,
and Ir (Kennicutt \cite{ken92}). The template producing the highest
value of $\cal R$, i.e., the parameter given by RVSAO and related to
the signal-to-noise ratio of the correlation peak, was
chosen. Moreover, all spectra and their best correlation functions
were examined visually to verify the redshift determination.

In eleven cases (IDs.~1, 6, 14, 31, 48, 49, 89, 110, 113, 122 and 131;
see Table~\ref{catalogA1758}), we had to rely on the EMSAO package to
obtain an estimate of the redshift.

The formal errors as given by the cross-correlation are known to be
smaller than the true errors (e.g., Malumuth et al. \cite{mal92};
Bardelli et al. \cite{bar94}; Ellingson \& Yee \cite{ell94}; Quintana
et al. \cite{qui00}). Duplicate observations for the same galaxy
allowed us to estimate the true intrinsic errors in data of the same
quality taken with the same instrument (e.g. Barrena et
al. \cite{bar09}).  Here we have double determinations for six
galaxies, therefore we decided to apply the procedure already applied
in Barrena et al. (\cite{bar09}), obtaining that true errors are
larger than formal cross-correlation errors by a factor of 2.2. For
the galaxies with two redshift estimates, we used the mean of the two
measurements and the corresponding errors. As for the radial
velocities estimated through EMSAO we assumed the largest between the
formal error and 100 \kss.

Our spectroscopic catalog lists 137 galaxies in the field of A1758N
(see Fig.~\ref{figottico}). The median error in $cz$ is 76 \kss.

Out of 137 galaxies, only four have a previously measured redshift as
listed by NED. In particular, we looked at SDSS spectroscopic data
(Data Release 7) finding eight ``likely'' member galaxies with
measured spectroscopic redshift in the field we sampled with the
TNG. Out of these, only three galaxies are in common with our
sample. The three redshifts agree with our estimates within
1-2$\sigma$ and the difference ${\rm v_{\rm our}}-{\rm v_{\rm SDSS}}$
ranges between 72--92 \ks for all three galaxies. Owing to the small
improvement we decided to not add the other five, preferring to work
with a homogeneous sample.  However, in our discussions we also
consider the BCG of A1758S
(R.A.=$13^{\mathrm{h}}32^{\mathrm{m}}32\dotsec96$ and Dec.=$+50\degree
25\arcmm 02.5\arcs$-- J2000.0, $cz=81802\pm52$, $r^\prime$=17.13;
hereafter BCGS).

We used public photometric data from the SDSS (Data Release 7). In
particular, we used $g^\prime$, $r^\prime$, and $i^\prime$ magnitudes,
already corrected for the Galactic extinction, and considered galaxies
within a radius of 25\arcm from the cluster center.

Table~\ref{catalogA1758} lists the velocity catalog (see also
Fig.~\ref{figottico}): identification number of each galaxy and
members, ID and IDm (Cols.~1 and 2, respectively); right ascension and
declination, $\alpha$ and $\delta$ (J2000, Col.~3); $r^\prime$ SDSS
magnitude (Col.~4); heliocentric radial velocities, ${\rm
  v}=cz_{\sun}$ (Col.~5) with errors, $\Delta {\rm v}$ (Col.~6).

We also used public photometric data from the CFHT archive taken with
the Megaprime/Megacam camera. We retrieved from the CADC Megapipe
archive (Gwyn \cite{gwy09}) the catalogs for the images in the $g^{\rm
  Mega}$ and $r^{\rm Mega}$ bands\footnote{For a comparison between
  Megacam and SDSS filters, check the web page
  http://www2.cadc-ccda.hia-iha.nrc-cnrc.gc.ca/megapipe/docs/filters.html}
and corrected the magnitudes for the Galactic extinction. The total
area covered by the images was $1.05\times1.16$ deg$^2$. The estimated
limiting magnitudes (at the 5$\sigma$ c.l.) are $g^{\rm Mega}=27.0$ and
$r^{\rm Mega}=26.8$.

%%new commands
%\def\lesssim{\mathrel{\hbox{\rlap{\hbox{\lower4pt\hbox{$\sim$}}}\hbox{$<$}}}}
%\def\gtrsim{\mathrel{\hbox{\rlap{\hbox{\lower4pt\hbox{$\sim$}}}\hbox{$>$}}}}
%\newcommand{\mincir}{\raise -2.truept\hbox{\rlap{\hbox{$\sim$}}\raise5.truept
%\hbox{$<$}\ }}
%\newcommand{\magcir}{\raise -2.truept\hbox{\rlap{\hbox{$\sim$}}\raise5.truept
%\hbox{$>$}\ }}
%\newcommand{\siml}{\raise -2.truept\hbox{\rlap{\hbox{$\sim$}}\raise5.truept
%\hbox{$<$}\ }}
%\newcommand{\simg}{\raise -2.truept\hbox{\rlap{\hbox{$\sim$}}\raise5.truept
%\hbox{$>$}\ }}
%\newcommand{\be}{\begin{equation}}
%\newcommand{\ee}{\end{equation}}
%\newcommand{\ba}{\begin{eqnarray}}
%\newcommand{\ea}{\end{eqnarray}}
%\newcommand {\h} {$h^{-1}$ Mpc $ \;$}
%\newcommand {\kpc} {$h^{-1}$ kpc}
%\newcommand {\hh} {$h^{-1}$ Mpc}
%\newcommand {\ks} {km~s$^{-1} \;$}
%\newcommand {\kss} {km~s$^{-1}$}
%\newcommand {\mpc} {$Mpc \;$}
%\newcommand {\msun} {$h^{-1} \  M_{\odot} \;$}
%\newcommand {\m} {$M_{\odot} \;$}
%\newcommand {\ml} {$h \, M_{\odot}/L_{\odot} \;$}
%\newcommand {\mll} {$h \, M_{\odot}/L_{\odot}$}
%\newcommand{\vel}{\,{\rm km\,s^{-1}}}
%\newcommand{\tng}{\mathrm{T}}
%\newcommand{\sds}{\mathrm{S}}
%\newcommand{\tns}{\mathrm{T+S}}
%%
%\addtocounter{table}{-2}
\begin{table}[!ht]
        \caption[]{Velocity catalog of 137 spectroscopically measured
          galaxies in the field of the cluster A1758N. $\dagger$ and
          $\ddagger$ highlight the IDs of the BCGN(NW) and BCGN(SE),
          respectively.}
         \label{catalogA1758}
              $$ 
        % \begin{array}{p{0.5\linewidth}l}
           \begin{array}{r r c c r r}
            \hline
            \noalign{\smallskip}
            \hline
            \noalign{\smallskip}

\mathrm{ID} & \mathrm{IDm} &\mathrm{\alpha},\mathrm{\delta}\,(\mathrm{J}2000)  & r^{\prime}& \mathrm{v}\,\,\,\,\,&\mathrm{\Delta}\mathrm{v}\\
  & & & &\mathrm{(\,km}&\mathrm{s^{-1}\,)}\\
            \hline
            \noalign{\smallskip}  

 1           & -         &13\ 32\ 13.95  ,+50\ 36\ 23.2 &21.75     &125775  & 71  \\
 2           & -         &13\ 32\ 14.70  ,+50\ 34\ 52.4 &19.65     & 99042  & 82  \\ 
 3           & -         &13\ 32\ 15.38  ,+50\ 35\ 18.1 &20.01     & 85168  &104  \\ 
 4           & 1         &13\ 32\ 16.88  ,+50\ 33\ 19.3 &19.79     & 80993  & 94  \\ 
 5           & -         &13\ 32\ 17.62  ,+50\ 37\ 27.2 &19.11     & 41801  &173  \\ 
 6           & -         &13\ 32\ 17.85  ,+50\ 33\ 39.6 &20.10     & 32788  &100  \\ 
 7           & -         &13\ 32\ 17.92  ,+50\ 37\ 39.5 &20.18     & 88373  &128  \\ 
 8           & -         &13\ 32\ 18.32  ,+50\ 36\ 47.6 &20.23     & 98217  & 93  \\ 
 9           & 2         &13\ 32\ 18.98  ,+50\ 36\ 04.4 &21.70     & 84170  &129  \\ 
10           & -         &13\ 32\ 19.28  ,+50\ 36\ 03.8 &20.21     & 85959  &128  \\ 
11           & -         &13\ 32\ 19.45  ,+50\ 34\ 11.1 &20.31     &112194  & 60  \\ 
12           & -         &13\ 32\ 22.27  ,+50\ 35\ 08.9 &19.89     & 98001  & 50  \\ 
13           & -         &13\ 32\ 22.42  ,+50\ 34\ 48.9 &19.90     & 97790  & 69  \\ 
14           & -         &13\ 32\ 22.47  ,+50\ 37\ 04.5 &21.95     & 97872  &100  \\ 
15           & -         &13\ 32\ 22.59  ,+50\ 35\ 27.4 &20.17     & 98137  & 88  \\ 
16           & 3         &13\ 32\ 22.89  ,+50\ 36\ 04.7 &20.36     & 84969  & 60  \\ 
17           & 4         &13\ 32\ 25.27  ,+50\ 34\ 03.5 &18.32     & 82720  & 48  \\ 
18           & -         &13\ 32\ 25.98  ,+50\ 38\ 18.2 &20.16     & 85091  &114  \\ 
19           & 5         &13\ 32\ 27.59  ,+50\ 33\ 31.4 &20.41     & 84801  & 98  \\ 
20           & 6         &13\ 32\ 27.59  ,+50\ 37\ 44.5 &19.84     & 82907  & 92  \\ 
21           & 7         &13\ 32\ 29.23  ,+50\ 34\ 10.0 &19.73     & 83370  & 92  \\ 
22           & -         &13\ 32\ 29.57  ,+50\ 35\ 25.8 &20.33     & 74891  &116  \\ 
23           & 8         &13\ 32\ 29.93  ,+50\ 37\ 21.5 &20.00     & 82044  & 54  \\ 
24           & -         &13\ 32\ 30.23  ,+50\ 36\ 37.1 &20.32     &113677  & 79  \\ 
25           & 9         &13\ 32\ 30.27  ,+50\ 33\ 49.0 &19.89     & 82680  & 70  \\ 
26           &10         &13\ 32\ 30.62  ,+50\ 36\ 25.5 &20.04     & 82111  & 66  \\ 
27           & -         &13\ 32\ 31.64  ,+50\ 36\ 17.7 &19.73     & 98544  & 76  \\ 
28           &11         &13\ 32\ 31.88  ,+50\ 34\ 49.9 &20.37     & 83238  &123  \\ 
29           & -         &13\ 32\ 32.64  ,+50\ 34\ 27.9 &18.55     & 53251  & 53  \\ 
30           &12         &13\ 32\ 32.86  ,+50\ 36\ 04.5 &19.96     & 85335  & 62  \\ 
31           & -         &13\ 32\ 33.31  ,+50\ 34\ 02.5 &21.70     &187664  &100  \\ 
32           & -         &13\ 32\ 33.72  ,+50\ 31\ 03.4 &19.90     &112311  & 98  \\ 
33           & -         &13\ 32\ 33.86  ,+50\ 30\ 50.0 &19.01     &112310  & 75  \\ 
34           & -         &13\ 32\ 34.12  ,+50\ 30\ 39.7 &20.22     &113091  &116  \\ 
35           &13         &13\ 32\ 34.31  ,+50\ 33\ 03.1 &19.47     & 87495  & 50  \\ 
36           &14         &13\ 32\ 34.41  ,+50\ 32\ 11.3 &19.78     & 80234  & 41  \\ 
37           &15         &13\ 32\ 34.49  ,+50\ 33\ 18.5 &19.11     & 84766  & 70  \\ 
38           &16         &13\ 32\ 34.57  ,+50\ 33\ 16.2 &21.48     & 83263  &223  \\ 
39           &17         &13\ 32\ 34.69  ,+50\ 31\ 39.1 &20.45     & 84767  & 92  \\ 
40           &18         &13\ 32\ 34.70  ,+50\ 33\ 24.7 &19.43     & 81303  & 72  \\ 
41           &19         &13\ 32\ 34.75  ,+50\ 31\ 43.3 &20.46     & 86327  & 79  \\ 
42           &20         &13\ 32\ 34.96  ,+50\ 32\ 37.3 &18.75     & 84194  & 44  \\   
43          &  -          &13\ 32\ 35.07  ,+50 \ 38\ 30.5 &19.68     & 98294   &119   \\
44          & 21          &13\ 32\ 35.21  ,+50 \ 32\ 36.2 &18.25     & 81900   & 71   \\
45          & 22          &13\ 32\ 35.69  ,+50 \ 33\ 38.1 &20.29     & 83293   & 54   \\                            
                                       \noalign{\smallskip}			    
            \hline					    
            \noalign{\smallskip}			    
            \hline					    
         \end{array}
     $$        
         \end{table}
\addtocounter{table}{-1}
\begin{table}[!ht]
          \caption[ ]{Continued.}
     $$        
           \begin{array}{r r c c r r}
            \hline
            \noalign{\smallskip}
            \hline
            \noalign{\smallskip}
               
\mathrm{ID} & \mathrm{IDm} &\mathrm{\alpha},\mathrm{\delta}\,(\mathrm{J}2000)  & r^{\prime}& \mathrm{v}\,\,\,\,\,&\mathrm{\Delta}\mathrm{v}\\
  & & & &\mathrm{(\,km}&\mathrm{s^{-1}\,)}\\
               
            \hline
            \noalign{\smallskip}
                              
46          &  -          &13\ 32\ 35.73  ,+50 \ 34\ 49.8 &20.83     &113693   & 82   \\
47          & 23          &13\ 32\ 35.84  ,+50 \ 33\ 52.0 &20.50     & 83709   & 76   \\
48          &  -          &13\ 32\ 36.46  ,+50 \ 35\ 50.6 &20.43     & 87368   &100   \\
49          &  -          &13\ 32\ 36.88  ,+50 \ 30\ 37.7 &20.50     &112681   &100   \\
50          & 24          &13\ 32\ 36.89  ,+50 \ 32\ 59.7 &19.54     & 83963   & 63   \\
51          & 25          &13\ 32\ 37.58  ,+50 \ 33\ 05.7 &19.35     & 82279   & 44   \\
52          & 26          &13\ 32\ 38.04  ,+50 \ 30\ 57.2 &20.58     & 84344   & 79   \\
53          & 27          &13\ 32\ 38.33  ,+50 \ 31\ 30.3 &20.92     & 86862   &255   \\
54          & 28          &13\ 32\ 38.33  ,+50 \ 33\ 30.1 &20.48     & 84176   &120   \\
55          &\dagger29    &13\ 32\ 38.41  ,+50 \ 33\ 35.7 &17.15     & 83489   & 57   \\
56          & 30          &13\ 32\ 38.46  ,+50 \ 31\ 41.4 &20.59     & 85204   &113   \\
57          & 31          &13\ 32\ 38.48  ,+50 \ 32\ 53.0 &20.14     & 83376   & 76   \\
58          & 32          &13\ 32\ 38.58  ,+50 \ 33\ 43.9 &19.52     & 83791   & 79   \\
59          & 33          &13\ 32\ 39.43  ,+50 \ 34\ 45.0 &18.97     & 83208   & 50   \\
60          & 34          &13\ 32\ 39.54  ,+50 \ 33\ 45.0 &20.13     & 83880   & 76   \\
61          & 35          &13\ 32\ 39.54  ,+50 \ 34\ 32.0 &19.01     & 87769   & 85   \\
62          & 36          &13\ 32\ 39.76  ,+50 \ 33\ 14.4 &20.40     & 82872   &128   \\
63          & 37          &13\ 32\ 39.80  ,+50 \ 32\ 41.0 &19.37     & 83803   & 84   \\
64          & 38          &13\ 32\ 39.94  ,+50 \ 32\ 23.9 &20.94     & 83497   &172   \\
65          & 39          &13\ 32\ 40.35  ,+50 \ 34\ 49.4 &19.50     & 85818   & 70   \\
66          & 40          &13\ 32\ 40.51  ,+50 \ 35\ 39.6 &18.56     & 81400   & 47   \\
67          & 41          &13\ 32\ 40.52  ,+50 \ 32\ 16.0 &21.97     & 85932   &182   \\
68          & 42          &13\ 32\ 40.59  ,+50 \ 33\ 15.4 &19.63     & 80255   & 54   \\
69          & 43          &13\ 32\ 40.96  ,+50 \ 33\ 46.3 &17.93     & 83667   & 38   \\
70          & 44          &13\ 32\ 41.47  ,+50 \ 32\ 10.7 &20.38     & 80544   & 57   \\
71          & 45          &13\ 32\ 42.03  ,+50 \ 34\ 34.7 &20.19     & 87429   &132   \\
72          & 46          &13\ 32\ 42.21  ,+50 \ 32\ 30.1 &20.56     & 82939   & 85   \\
73          & 47          &13\ 32\ 42.35  ,+50 \ 32\ 33.1 &20.94     & 81072   & 91   \\
74          & 48          &13\ 32\ 43.30  ,+50 \ 29\ 26.2 &18.99     & 82769   & 63   \\
75          & 49          &13\ 32\ 43.43  ,+50 \ 33\ 28.8 &18.80     & 85478   & 60   \\
76          & 50          &13\ 32\ 43.77  ,+50 \ 31\ 14.6 &19.35     & 83504   & 50   \\
77          & 51          &13\ 32\ 44.77  ,+50 \ 33\ 26.0 &19.78     & 84664   & 97   \\
78          & 52          &13\ 32\ 44.96  ,+50 \ 31\ 57.2 &20.03     & 84337   & 82   \\
79          & 53          &13\ 32\ 45.35  ,+50 \ 33\ 24.6 &19.47     & 85798   & 82   \\
80          & 54          &13\ 32\ 45.94  ,+50 \ 32\ 04.7 &19.50     & 79995   & 76   \\
81          & 55          &13\ 32\ 46.36  ,+50 \ 36\ 41.5 &18.88     & 86035   & 47   \\
82          & 56          &13\ 32\ 46.90  ,+50 \ 31\ 48.5 &20.48     & 83825   &113   \\
83          & 57          &13\ 32\ 47.00  ,+50 \ 32\ 02.0 &19.77     & 84249   & 76   \\
84          & 58          &13\ 32\ 47.92  ,+50 \ 32\ 09.6 &20.03     & 82909   & 82   \\        
 85           & -          &13\ 32\ 48.47  ,+50\ 28\ 06.7 &19.29     &109410  & 85   \\  
 86           &59          &13\ 32\ 48.71  ,+50\ 31\ 54.9 &19.94     & 82757  & 63   \\  
 87           &60          &13\ 32\ 48.72  ,+50\ 31\ 21.7 &19.47     & 83562  & 63   \\  
 88           &61          &13\ 32\ 49.25  ,+50\ 31\ 31.1 &19.89     & 82110  & 88   \\  
 89           & -          &13\ 32\ 49.28  ,+50\ 28\ 50.3 &20.45     &161829  &100   \\  
 90           &62          &13\ 32\ 49.32  ,+50\ 33\ 55.9 &19.95     & 79788  & 76   \\  
                              
                                       \noalign{\smallskip}			    
            \hline					    
            \noalign{\smallskip}			    
            \hline					    
         \end{array}
     $$        
         \end{table}
\addtocounter{table}{-1}
\begin{table}[!ht]
          \caption[ ]{Continued.}
     $$        
           \begin{array}{r r c c r r}
            \hline
            \noalign{\smallskip}
            \hline
            \noalign{\smallskip}
               
\mathrm{ID} & \mathrm{IDm} &\mathrm{\alpha},\mathrm{\delta}\,(\mathrm{J}2000)  & r^{\prime}& \mathrm{v}\,\,\,\,\,&\mathrm{\Delta}\mathrm{v}\\
  & & & &\mathrm{(\,km}&\mathrm{s^{-1}\,)}\\

            \hline
            \noalign{\smallskip}
                              
 91           & -          &13\ 32\ 49.73  ,+50\ 28\ 41.6 &18.64     & 55954  & 60   \\  
 92           &63          &13\ 32\ 50.00  ,+50\ 32\ 26.0 &19.98     & 84288  & 79   \\  
 93           &64          &13\ 32\ 50.03  ,+50\ 33\ 53.8 &20.40     & 81273  & 60   \\  
 94           & -          &13\ 32\ 50.62  ,+50\ 27\ 52.3 &19.49     & 80403  & 72   \\  
 95           &65          &13\ 32\ 50.72  ,+50\ 27\ 46.7 &19.55     & 82691  & 79   \\  
 96           &66          &13\ 32\ 51.01  ,+50\ 33\ 08.8 &19.89     & 84526  & 79   \\  
 97           &67          &13\ 32\ 51.10  ,+50\ 27\ 45.8 &19.70     & 81557  & 82   \\  
 98           &68          &13\ 32\ 51.42  ,+50\ 33\ 04.7 &17.95     & 85536  & 57   \\  
 99           & -          &13\ 32\ 51.50  ,+50\ 33\ 10.3 &20.16     &130803  &148   \\  
100           &69          &13\ 32\ 51.79  ,+50\ 27\ 13.5 &18.01     & 83197  & 54   \\  
101           &\ddagger70  &13\ 32\ 52.10  ,+50\ 31\ 34.1 &17.02     & 83864  & 70   \\  
102           &71          &13\ 32\ 52.21  ,+50\ 31\ 21.8 &20.36     & 83139  & 72   \\
103           & -          &13\ 32\ 52.33  ,+50\ 27\ 42.8 &20.60     &109756  &204   \\  
104           &72          &13\ 32\ 52.41  ,+50\ 31\ 12.8 &20.23     & 82717  & 75   \\  
105           & -          &13\ 32\ 52.93  ,+50\ 31\ 46.1 &18.52     & 79682  & 41   \\  
106           &73          &13\ 32\ 53.10  ,+50\ 31\ 35.5 &19.64     & 83648  & 63   \\  
107           & -          &13\ 32\ 53.24  ,+50\ 30\ 29.0 &18.69     & 99000  & 47   \\  
108           &74          &13\ 32\ 53.57  ,+50\ 27\ 36.0 &20.42     & 83346  & 94   \\  
109           & -          &13\ 32\ 53.69  ,+50\ 30\ 53.3 &19.06     & 98877  &104   \\  
110           & -          &13\ 32\ 53.75  ,+50\ 26\ 56.2 &20.81     &212245  &100   \\  
111           &75          &13\ 32\ 53.92  ,+50\ 29\ 57.7 &19.90     & 82041  & 66   \\  
112           &76          &13\ 32\ 53.94  ,+50\ 32\ 20.7 &19.13     & 79836  & 47   \\
113           & -          &13\ 32\ 54.86  ,+50\ 30\ 27.8 &20.89     & 52848  &100   \\  
114           &77          &13\ 32\ 54.94  ,+50\ 33\ 14.7 &19.82     & 79373  & 54   \\  
115           &78          &13\ 32\ 55.00  ,+50\ 31\ 27.6 &21.38     & 84114  &154   \\  
116           &79          &13\ 32\ 55.15  ,+50\ 31\ 25.4 &18.85     & 85175  & 72   \\  
117           &80          &13\ 32\ 55.61  ,+50\ 31\ 28.5 &19.01     & 85239  & 47   \\
118           & -          &13\ 32\ 55.64  ,+50\ 30\ 55.8 &18.83     & 55708  & 98   \\  
119           & -          &13\ 32\ 56.08  ,+50\ 30\ 17.5 &18.98     & 31474  & 70   \\  
120           &81          &13\ 32\ 59.32  ,+50\ 30\ 19.9 &20.66     & 83353  &230   \\  
121           &82          &13\ 32\ 59.59  ,+50\ 29\ 27.6 &19.24     & 82055  & 50   \\  
122           & -          &13\ 32\ 59.75  ,+50\ 30\ 05.4 &20.61     & 86250  &100   \\  
123           &83          &13\ 33\ 02.05  ,+50\ 27\ 17.1 &19.84     & 82098  & 54   \\  
124           &84          &13\ 33\ 02.07  ,+50\ 29\ 28.0 &17.92     & 83651  & 47   \\  
125           &85          &13\ 33\ 02.12  ,+50\ 30\ 09.2 &20.88     & 83434  & 76   \\  
126           & -          &13\ 33\ 02.71  ,+50\ 31\ 42.3 &18.87     & 80377  & 82   \\  
127           & -          &13\ 33\ 03.67  ,+50\ 28\ 25.3 &19.87     & 93943  & 91   \\  
128           &86          &13\ 33\ 06.72  ,+50\ 31\ 24.1 &18.62     & 85598  & 47   \\        
129           & -          &13\ 33\ 07.43  ,+50\ 29\ 19.6 &20.35     & 82058  & 62   \\        
130           &87          &13\ 33\ 07.48  ,+50\ 32\ 00.7 &18.93     & 82508  & 47   \\        
131           & -          &13\ 33\ 09.83  ,+50\ 28\ 33.2 &19.57     & 87007  &100   \\        
132           &88          &13\ 33\ 09.90  ,+50\ 29\ 02.3 &19.94     & 83139  & 76   \\        
133           &89          &13\ 33\ 10.19  ,+50\ 28\ 51.3 &18.92     & 82496  & 72   \\        
134           & -          &13\ 33\ 10.72  ,+50\ 31\ 15.6 &18.87     & 86227  & 40   \\        
135           &90          &13\ 33\ 10.93  ,+50\ 28\ 53.7 &19.90     & 82857  & 91   \\        
136           &91          &13\ 33\ 12.99  ,+50\ 29\ 01.3 &18.52     & 82798  & 41   \\        
137           &92          &13\ 33\ 13.84  ,+50\ 29\ 55.3 &20.77     & 83782  & 94   \\        

                        \noalign{\smallskip}			    
            \hline					    
            \noalign{\smallskip}			    
            \hline					    
         \end{array}
     $$ 
\end{table}

No evident unique dominant galaxy is present in A1758N, instead there
are two BCGs: one in the NW subcluster (ID.~55, $r^{\prime}=17.15$,
hereafter BCGN(NW)) and one in the SE subcluster (ID.~101,
$r^{\prime}=17.02$, hereafter BCGN(SE)). They are separated by
$\sim0.85$ \h and are $\sim0.8$ mags brighter than other bright
galaxies in our sample. Moreover, Durret et al. (\cite{dur11}) found
that they are characterized by a surface brightness profile flatter
than those of other bright galaxies and suggested that the two BCGs
are really dominant galaxies in their respective subcluster.

\subsection{Radio and X-ray emitting galaxies in the field of A1758N}

Our spectroscopic catalog lists some radio galaxies observed in the
sky region of A1758N. At $\sim$14\arcs NE of BCGN(SE) we identify the
galaxy ID~105 with the radio source 1330+507 (O'Dea \& Owen
\cite{ode85}, see our Fig.~\ref{figimage}), one of the most powerful
narrow-angle tail (NAT) radio sources known with the tail pointing to
the SE. Rizza et al. (\cite{riz03}) used high-resolution VLA data to
list a total of 11 radio sources in the field of A1758, some of them
members of A1758N. Interestingly, source \#1 of Rizza et al. (see
their Table~3) is our ID~55 (the BCGN(NW)). Other radio sources are
ID~51 (\#2 of Rizza et al.), ID~33 (\#3, background galaxy), ID~124
(\#6, the third-brightest galaxy in our spectroscopic sample) and
ID~136 (\#10).

According to Hart et al. (\cite{har09}), BCGN(NW) is also a pointlike
X-ray source. A look at the original (non-smoothed) {\em Chandra}
X-ray image reveals that ID~124 is a faint X-ray source, too.

\section{Analysis of the spectroscopic sample}
\label{ana}

\subsection{Member selection}
\label{memb}

\begin{figure}
\centering
\resizebox{\hsize}{!}{\includegraphics{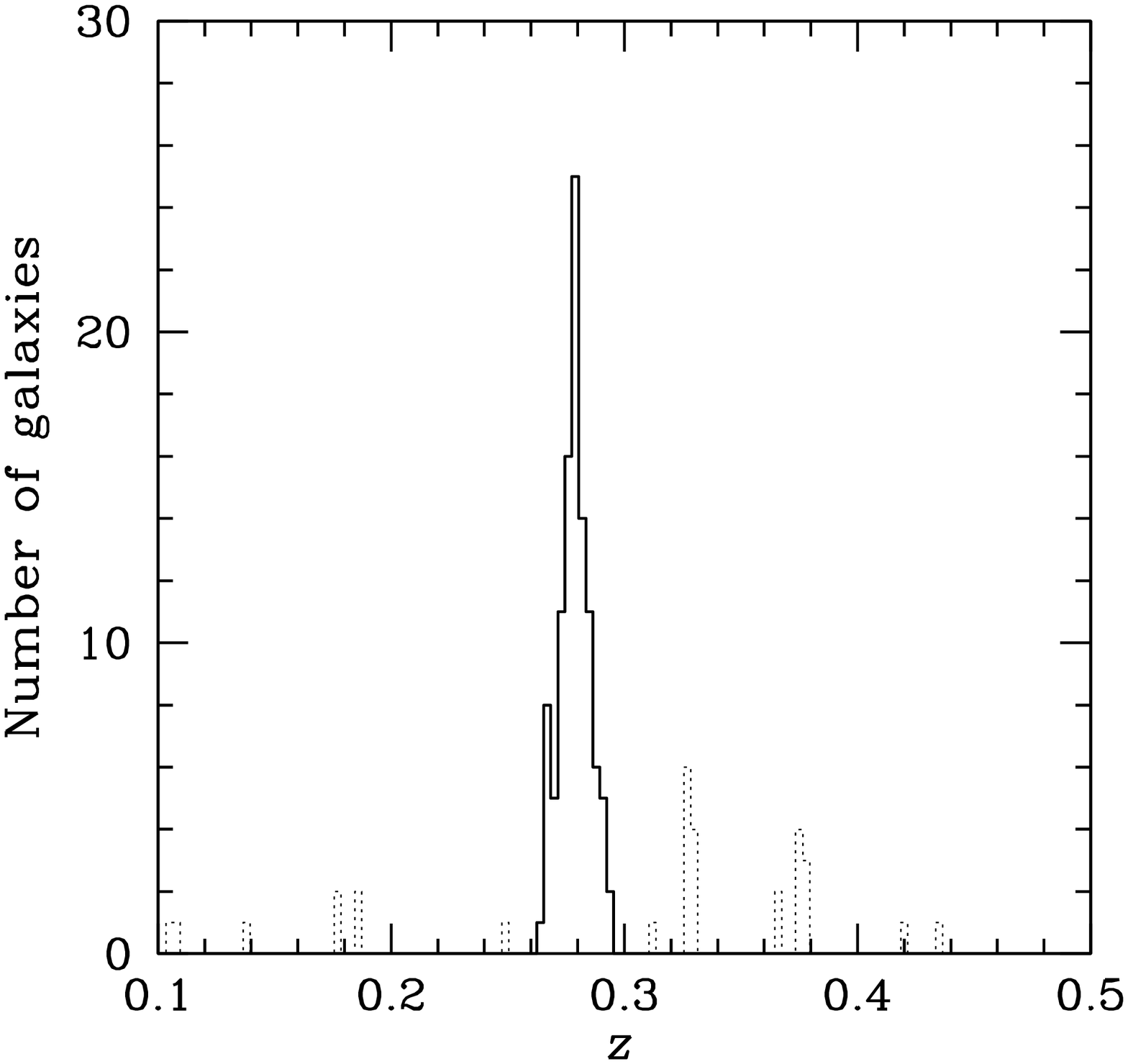}}
\caption
{Redshift galaxy distribution. The solid line histogram refers to the
  104 galaxies assigned to A1758N according to the DEDICA
  reconstruction method.}
\label{fighisto}
\end{figure}

To select cluster members among the 137 galaxies with redshifts, we
followed a two-step procedure. We first used the 1D adaptive-kernel
method (hereafter DEDICA, Pisani \cite{pis93} and \cite{pis96}; see
also Fadda et al. \cite{fad96}; Girardi et al. \cite{gir96}). We
searched for significant peaks in the velocity distribution at $>$99\%
c.l.. This procedure detects A1758N as a peak at $z\sim0.278$
populated by 104 galaxies considered as candidate cluster members (in
the range $79\,373\leq {\rm v} \leq 88\,373$ \kss, see
Fig.~\ref{fighisto}). Out of 33 nonmembers, 8/25 are
foreground/background galaxies.

\begin{figure}
\centering 
\resizebox{\hsize}{!}{\includegraphics{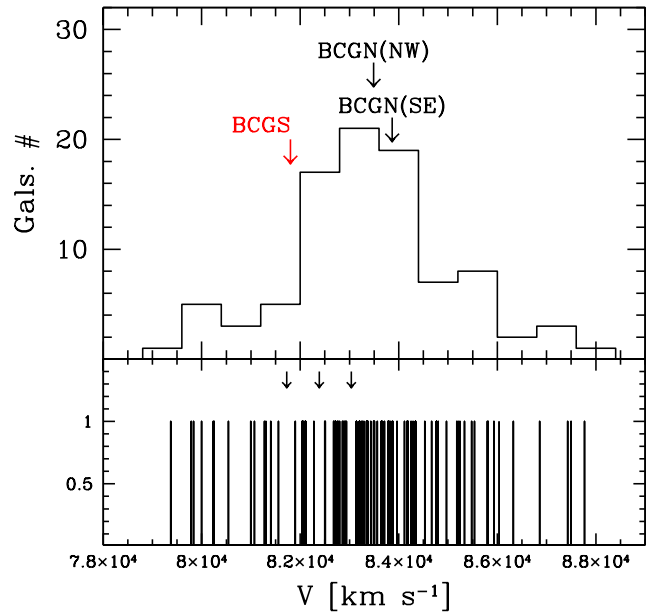}}
\caption
{The 92 galaxies assigned to the cluster A1758N.  {\em Upper panel}:
  Galaxy velocity distribution. The arrows indicate the velocities of the
  BCGN(NW) and BCGN(SE). We also show the velocity of BCGS of A1758S
  as taken from the SDSS. {\em Lower panel}: Stripe density plot where the
arrows indicate the positions of the significant gaps.
}
\label{figstrip}
\end{figure}

All galaxies assigned to the cluster peak were analyzed in the second
step, which combines position and velocity information, i.e., the
``shifting gapper'' method by Fadda et al. (\cite{fad96}).  This
procedure rejects galaxies that are too far in velocity from the main
body of galaxies within a fixed bin that shifts along the distance
from the cluster center.  The procedure is iterated until the number
of cluster members converges on a stable value.  Following Fadda et
al. (\cite{fad96}), we used a gap of $1000$ \ks -- in the cluster
rest-frame -- and a bin of 0.6 \hh, or large enough to include 15
galaxies. For the center of A1758N we adopted the position of the
BCGN(NW) [R.A.=$13^{\mathrm{h}}32^{\mathrm{m}}38\dotsec41$,
  Dec.=$+50\degree 33\arcmm 35.7\arcs$(J2000.0)]. We selected this
center because, contrary to the SE subcluster, the peak position of
the X-ray NW subclump coincides with that of the BCGN(NW). Moreover,
the NW subcluster is more luminous, massive, and hot than the SE
subcluster (Okabe et al. \cite{oka08}; Durret et al. \cite{dur11}),
which suggests that it is the primary component of A1758N.  The
``shifting gapper'' procedure rejected another 12 interlopers. Most of
them are not very far from the main body, but we are confident enough
in our rejection because 6 out of 12 are emission line galaxies (ELGs
hereafter), which increases their probability of belonging to the
field or to the infalling cluster region (see Sect.~\ref{stru} for a
more detailed discussion on these ELGs). Finally, we obtained a sample
of 92 fiducial cluster members (see Fig.~\ref{figstrip}). The removed
interlopers are shown in Fig.~\ref{figprof} (top panel).

\subsection{General cluster properties}
\label{prop}

By applying the biweight estimator to the 92 cluster members (Beers et
al. \cite{bee90}, ROSTAT software), we computed a mean cluster redshift
of $\left<z\right>=0.2782\pm$0.0005, i.e.
$\left<\rm{v}\right>=(83390\pm$139) \kss.  We estimated the LOS
velocity dispersion, $\sigma_{\rm V}$, by using the biweight estimator
and applying the cosmological correction and the standard correction
for velocity errors (Danese et al. \cite{dan80}).  We obtained
$\sigma_{\rm V}=1329_{-116}^{+135}$ \kss, where errors are estimated
through a bootstrap technique.

\begin{figure}
\centering
\resizebox{\hsize}{!}{\includegraphics{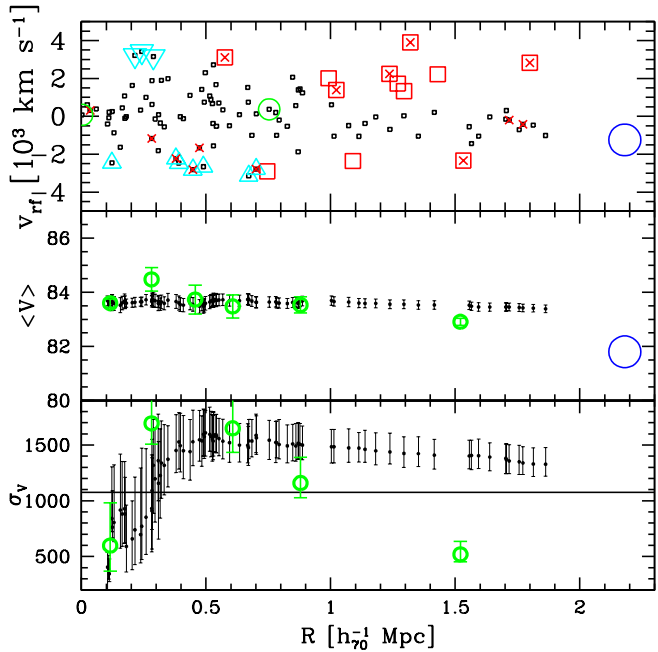}}
\caption
{{\em Top panel:} rest-frame velocity vs. projected distance from the
  cluster center (here the BCGN(NW)).  Small/black and large/red
  squares indicate the member galaxies and those rejected as
  interlopers by the shifting gapper procedure, respectively. Red
  crosses indicate ELGs.  The two green circles indicate the
  BCGN(NW) and the BCGN(SE).  The large/blue circle indicates the BCG
  of A1758S using the SDSS redshift.  The cyan triangles and rotated
  triangles indicate galaxies detected as members of HT2 and HT12
  subclusters detected through the Htree-method (see Sect.~\ref{3d}).
  {\em Middle and bottom panels:} differential (big circles) and
  integral (small points) profiles of mean velocity and LOS velocity
  dispersion, respectively.  For the differential profiles, we plot
  the values for six annuli from the center of the cluster, each
  containing 15 galaxies (large green symbols).  For the integral
  profiles, the mean and dispersion at a given (projected) radius from
  the cluster-center is estimated by considering all galaxies within
  that radius -- the first value computed on the five galaxies closest
  to the center. The error bands at the $68\%$ c.l. are also shown.
  In the lower panel, the horizontal line represents the X-ray
  temperature (7 keV) estimated from David \& Kempner (\cite{dav04})
  for the core of the two subclusters, see also Durret et
  al. (\cite{dur11}), transformed to $\sigma_{\rm V}$ assuming the
  density-energy equipartition between ICM and galaxies, i.e.
  $\beta_{\rm spec}=1$ (see Sect.~\ref{discu}).}
\label{figprof}
\end{figure}

To evaluate the robustness of the $\sigma_{\rm V}$ estimate, we
analyzed the velocity dispersion profile (Fig.~\ref{figprof}).  The
integral profile flattens in the external cluster regions, as found
for most nearby clusters (e.g., Fadda et al. \cite{fad96}; Girardi et
al. \cite{gir96}).  The sharp rising of the integral and
differential profiles in the internal region is discussed in
Sect.~\ref{stru}.

\subsection{Velocity distribution}
\label{1d}

We analyzed the velocity distribution to search for possible deviations
from Gaussianity that might provide important signatures of complex
dynamics. For the following tests, the null hypothesis is that the
velocity distribution is a single Gaussian.

We estimated three shape estimators, i.e., the kurtosis, the skewness,
and the scaled tail index STI.  We found STI=1.294, i.e. that the
velocity distribution departs from Gaussianity at the 95\%-99\%
c.l. (according to Table~1 of Bird \& Beers \cite{bir93}).  The
velocity distribution thus shows evidence for a heavy tailed
distribution.

We then investigated the gaps in the velocity distribution. We
followed the weighted gap analysis presented by Beers et
al. (\cite{bee91}; \cite{bee92}; ROSTAT software). We looked for
normalized gaps larger than 2.25, since in random draws of a Gaussian
distribution they arise at most in about $3\%$ of the cases,
independent of the sample size (Wainer and Schacht~\cite{wai78}). We
detected three significant gaps (at the $97\%$, $97\%$, and $98.6\%$
c.ls), which divide the cluster into four groups of 13, 9, 14, and 56
galaxies from low to high velocities (hereafter GV1, GV2, GV3, and
GV4, see Fig.~\ref{figstrip}). Both BCGN(NW) and BCGN(SE) were
assigned to the GV4 peak.

Following Ashman et al. (\cite{ash94}), we also applied the Kaye's
mixture model (KMM) algorithm. We found that a partition of 13, 7, 41,
and 31 galaxies is a significantly more accurate description of the
galaxy distribution than a single Gaussian (at the $\sim 97\%$
c.l.).

\subsection{Analysis of the 2D galaxy distribution}
\label{2d}

\begin{figure}
%\centering
\includegraphics[width=8cm]{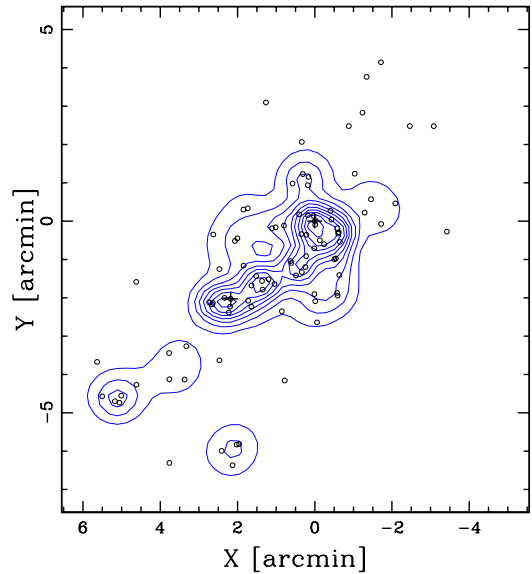}
\caption
{Spatial distribution on the sky and relative isodensity contour map
  of the 92 spectroscopic cluster members, obtained with the 2D-DEDICA
  method.  The BCGN(NW) is taken as the cluster center.  Crosses
  indicate the location of the BCGN(NW) and BCGN(SE).  }
\label{figk2z}
\end{figure}

By applying the 2D adaptive-kernel method (2D-DEDICA) to the positions
of the 92 member galaxies we found that the cluster is elongated in
the SE-NW direction. In particular, we detected three peaks with high
statistical significance: the main peak, which is coincident with the
BCGN(NW), a secondary peak coincident with the BCGN(SE), and a third
one, which is intermediate (Fig.~\ref{figk2z}).

Our spectroscopic data do not cover the entire cluster field and are
affected by magnitude incompleteness. More precisely, we found the
spectroscopic catalog has a $\sim 50$\% completeness for
$r^{\prime}<$19 and the completeness level quickly drops to $<$30\%
for $r^{\prime}>$21. To overcome these problems, we used the SDSS and
CFHT photometric data samples, which cover a larger spatial
region. The SDSS sample has the advantage to have photometry available
in several magnitude bands, while the CFHT sample allows us to extend
our analysis to fainter galaxies.

\begin{figure}
\centering
%\resizebox{\hsize}{!}{\includegraphics{figcm.eps}}
\includegraphics[width=8cm]{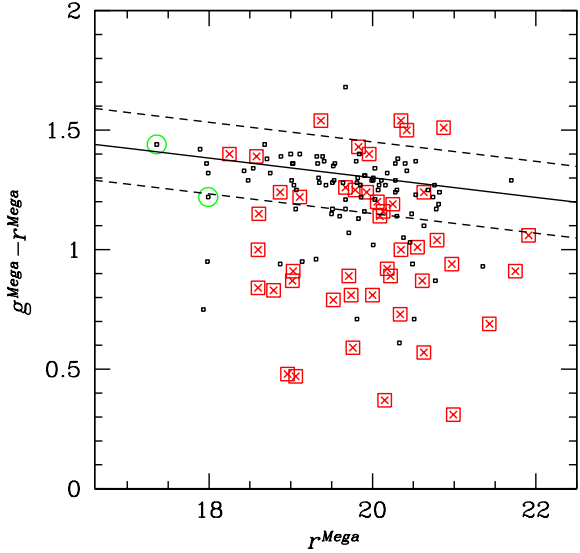}
\caption
{ CFHT $g^{\rm Mega}$--$r^{\rm Mega}$ vs. $r^{\rm Mega}$ diagram for
  galaxies with available spectroscopy.  Small/black squares indicate
  member galaxies and large/red squares indicate nonmember
  galaxies. Red crosses indicate ELGs.  The solid line gives the CMR
  determined on member galaxies; the dashed line is drawn at $\pm$0.15
  mag from this value.  The green circles indicate the BCGN(NW) and
  BCGN(SE) of A1758N: these are the only galaxies of the sample that
  likely suffer for photon count saturation in the CFHT catalog
  (because they are brighter than $r^{\rm Mega}\sim 17$).}
\label{figcm}
\end{figure}

From the CFHT photometric catalog we selected likely cluster members
on the basis of the ($g^{\rm Mega}$--$r^{\rm Mega}$) vs. $r^{\rm
  Mega}$ color-magnitude relation (hereafter CMR), which indicate the
locus of the red-sequence galaxies.  To determine the CMR we applied
the 2$\sigma$-clipping fitting procedure to the spectroscopic cluster
members. We obtained $g^{\rm Mega}$--$r^{\rm
  Mega}$=2.121--0.041$\times r^{\rm Mega}$ on 65 galaxies of our
spectroscopic sample, which agrees very well with the CMR directly
fitted on the photometric catalog recovered from CFHT images by Durret
et al. (\cite{dur11}).  Out of the CFHT photometric catalog, we
considered as likely ``red'' cluster members the objects lying within
0.15 mag of the CMR, where $\sim 0.15$ is the error associated to the
fitted intercept.  Figure~\ref{figcm} shows that our selection
criterion seems adequate to select red-sequence galaxies, which are
good tracers of the cluster substructure (e.g., Lubin et
al. \cite{lub00}) and, above all, avoiding nonmember galaxies that
might bias our 2D analysis.  As for the sample with $r^{\rm Mega}\leq
21$, we can check our selection of likely members: we can recover
$75\%$ of spectroscopic members (i.e. 68 out of 90), but $30\%$ of
nonmembers are also selected (13 out of 42).

\begin{figure}
%\centering
\includegraphics[width=8cm]{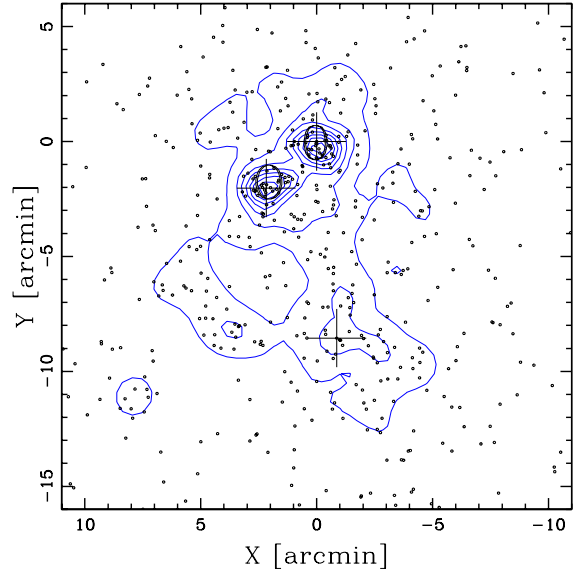}
\includegraphics[width=8cm]{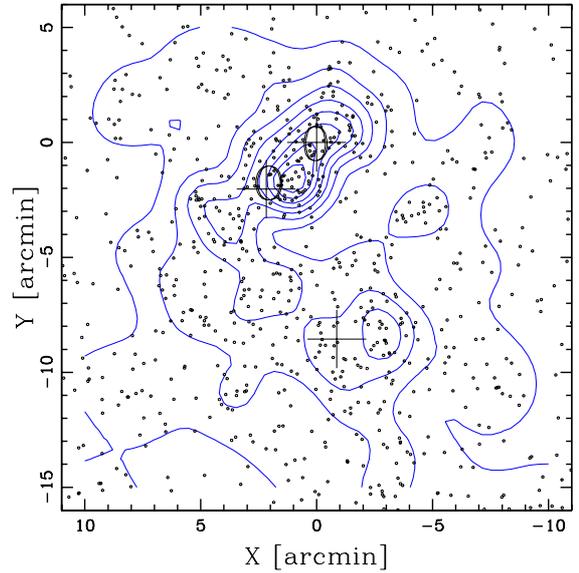}
\caption
{Spatial distribution on the sky and relative isodensity contour map
  of CFHT photometric cluster members with $r^{\rm Mega}\le 21$ ({\em
    upper panel}) and $21<r^{\rm Mega}\leq 23$ ({\em lower panel}),
  obtained with the 2D-DEDICA method.  The BCGN(NW) is taken as the
  cluster center.  Northern crosses indicate the location of the
  BCGN(NW) and BCGN(SE) of A1758N. The southern cross indicates the
  BCGS of A1758S. For A1758N, the two large ``O'' labels are
  centered on the two main peaks of the mass distribution (Okabe et
  al. \cite{oka08}).}
\label{figk2cfht}
\end{figure}

Figure~\ref{figk2cfht} shows the galaxy density contours obtained with
the 2D-DEDICA method for two samples: the CFHT likely members with
$r^{\rm Mega}\leq 21$ and those with $21<r^{\rm Mega}\leq 23$ (1657
and 3288 galaxies in the whole sample; 293 and 374 galaxies within 2
\hh, respectively).  As for the luminous galaxies, we found two very
significant peaks along the SE-NW direction, corresponding to the
BCGN(SE) and BCGN(NW): the NW-peak is the richest.  These two peaks
are separated by $\sim$0.75 \hh, with the SE-peak well coincident with
the BCGN(SE) and the NW-peak at a distance of $\sim$0.15 \h from the
BCGN(NW).  A third very significant peak lies in the South, coincident
with the position of the cluster A1758S.  As for the results with
$r^{\prime}\le 21$, Table~\ref{tabdedica2d} lists the number of
assigned members, $N_{\rm S}$ (Col.~2); the peak position (Col.~3);
the density (in arbitrary units relative to the densest peak fixed to
be 1), $\rho_{\rm S}$ (Col.~4); the value of $\chi^2$ for each peak,
$\chi ^2_{\rm S}$ (Col.~5).  The second sample (lower panel of
Fig.~\ref{figk2cfht}) contains galaxies fainter by two mags: as
discussed at the end of Sect.~\ref{stru}, these galaxies seem to trace
the whole A1758N system and not its substructure.

When more than two colors are available, it is very effective to
select cluster galaxies in the color-color space (Goto et al.
\cite{got02}) and both SDSS $r^{\prime}$--$i^{\prime}$ and
$g^{\prime}$--$r^{\prime}$ colors are generally used at $z<0.4$ (Lu et
al. \cite{lu09}; Lopes \cite{lop07}: his Eq.~1 and refs. therein).  In
particular, the difference between passive and star-forming galaxies
is larger in $g^{\prime}$--$r^{\prime}$ color than in
$r^{\prime}$--$i^{\prime}$ color (see Lu et al. \cite{lu09}, their
Fig.~6) but $r^{\prime}$--$i^{\prime}$ color has a much smaller
scatter (Lopes \cite{lop07}, his Fig.~1).  Out of the SDSS photometric
catalog, we considered as likely ``red'' cluster members the objects
lying within 0.1 and 0.2 mag of the CMRs ($r^{\prime}$--$i^{\prime}$
vs. $r^{\prime}$) and ($g^{\prime}$--$r^{\prime}$ vs. $r^{\prime}$),
respectively. When comparing this member selection to that performed
on the CFHT catalog, we obtain a comparable result with a modest,
better rejection of nonmembers. Indeed, we recover 67 out of 88
spectroscopic members and select only 10 out of 42 nonmembers (when
using the $g^{\prime}$--$r^{\prime}$ cut alone we obtain 67 out of 88
and 14 out of 42).  As for the galaxy density map, the SDSS result
confirms the result obtained with the bright CFHT sample: the NW-peak
is still richer than the SE-peak (77 vs. 52 galaxies), although the
SE-peak is now the densest (by a factor 1.5).

\begin{table}
        \caption[]{2D substructure from the CFHT photometric sample.}
         \label{tabdedica2d}
            $$
         \begin{array}{l r c c c }
            \hline
            \noalign{\smallskip}
            \hline
            \noalign{\smallskip}
\mathrm{Subclump} & N_{\rm S}^{\mathrm{\,\,\,\,a}} & \alpha({\rm J}2000),\,\delta({\rm J}2000)&\rho_{
\rm S}&\chi^2_{\rm S}\\
& & \mathrm{h:m:s,\degree:\arcmm:\arcs}&&\\
         \hline
         \noalign{\smallskip}
%\mathrm{2D-N(NW)\ (SDSS\ }r^{\prime}<21)    & 77&13\ 32\ 36.6+50\ 33\ 04&0.65&24\\
%\mathrm{2D-S    \ (SDSS\ }r^{\prime}<21)    & 52&13\ 32\ 24.9+50\ 24\ 41&0.28&13\\
%\mathrm{2D-N(SE)\ (SDSS\ }r^{\prime}<21)    & 38&13\ 32\ 52.3+50\ 31\ 36&1.00&35\\
\mathrm{2D-N(NW)\ (CFHT\ }r^{\rm Mega}<21)    & 87&13\ 32\ 37.7+50\ 33\ 19&1.00&54\\
\mathrm{2D-S    \ (CFHT\ }r^{\rm Mega}<21)    & 69&13\ 32\ 22.5+50\ 24\ 39&0.25&21\\
\mathrm{2D-N(SE)\ (CFHT\ }r^{\rm Mega}<21)    & 53&13\ 32\ 52.1+50\ 31\ 33&0.90&51\\
              \noalign{\smallskip}
              \noalign{\smallskip}
            \hline
            \noalign{\smallskip}
            \hline
         \end{array}
$$
\begin{list}{}{}  
\item[$^{\mathrm{a}}$] {Values are not background-subtracted.}
\end{list}
         \end{table}

\subsection{3D-analysis}
\label{3d}

The existence of correlations between positions and velocities of
cluster galaxies is a characteristic of true substructures. Here we
used several approaches to perform a 3D-analysis of the cluster.

We found no evidence for a significant velocity gradient (see, e.g.,
den Hartog \& Katgert \cite{den96} and Girardi et al. \cite{gir96}
for the details of the method).

We computed the $\Delta$-statistics devised by Dressler \& Schectman
(\cite{dre88}, hereafter DS-test), which is recommended by Pinkney et
al. (\cite{pin96}) as the most sensitive 3D test.  For each galaxy,
the deviation $\delta$ is defined as $\delta_i^2 =
[(N_{\rm{nn}}+1)/\sigma_{\rm{V}}^2][(\overline {\rm V_l} - \overline
  {\rm V})^2+(\sigma_{\rm V,l} - \sigma_{\rm V})^2]$, where the
subscript ``l'' denotes the local quantities computed over the
$N_{\rm{nn}}=10$ neighbors of the galaxy.  $\Delta$ is the sum of the
$\delta$ of the individual $N$ galaxies and gives the cumulative
deviation of the local kinematical parameters (mean velocity and
velocity dispersion) from the global cluster parameters.  The
significance of $\Delta$, i.e. of substructure, is checked by running
1000 Monte Carlo simulations, randomly shuffling the galaxy
velocities.  We found a significant presence of substructure at the
97\% c.l..

Following Pinkney et al. (\cite{pin96}; see also Ferrari et
al. \cite{fer03}), we applied two more classical 3D tests: the
$\epsilon$-test (Bird, \cite{bir93}) based on the projected mass
estimator and the centroid shift or $\alpha$-test (West \& Bothun
\cite{wes90}).  The details of these tests can be found in the above
papers. We merely point out that we considered ten as the number of
the nearest neighbors for each galaxy and used Monte Carlo simulations
to compute the substructure significance. The application of the
$\epsilon$-test leads to a significant presence of substructure at the
99.4\% c.l..

\begin{figure}
\centering 
\resizebox{\hsize}{!}{\includegraphics{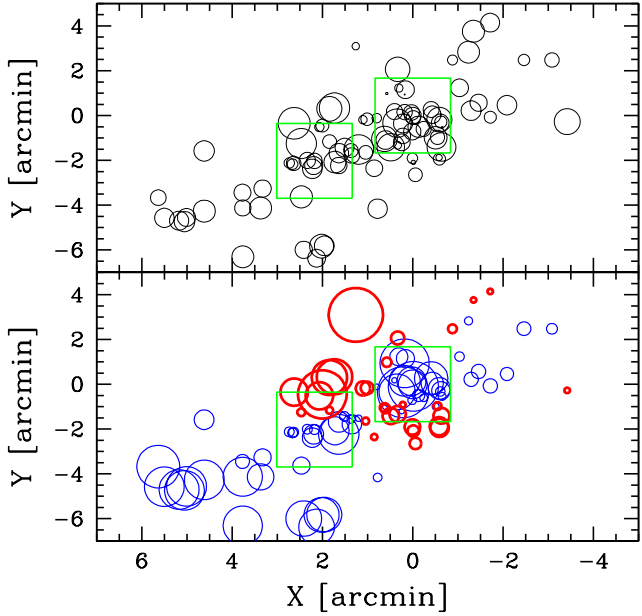}}
\caption
{ Spatial distribution of the 92 cluster members, each marked by a
  circle.  The BCGN(NW) is taken as the cluster center.  {\em Upper
    panel:} the cluster velocity field: the larger the circle, the
  larger the galaxy velocity.  {\em Lower panel:} the result of the
  (modified) DS-test: the larger the circle, the larger the deviation
  $\delta_{{\rm s},i}$ of the local velocity dispersion from the
  global velocity dispersion. Thin/blue and thick/red circles show
  where the local velocity dispersion is lower or higher than the
  global value.  The two large green squares indicate the positions of
  the two BCGs.  }
\label{figdssegno10}
\end{figure}

\begin{figure}
\centering 
\resizebox{\hsize}{!}{\includegraphics{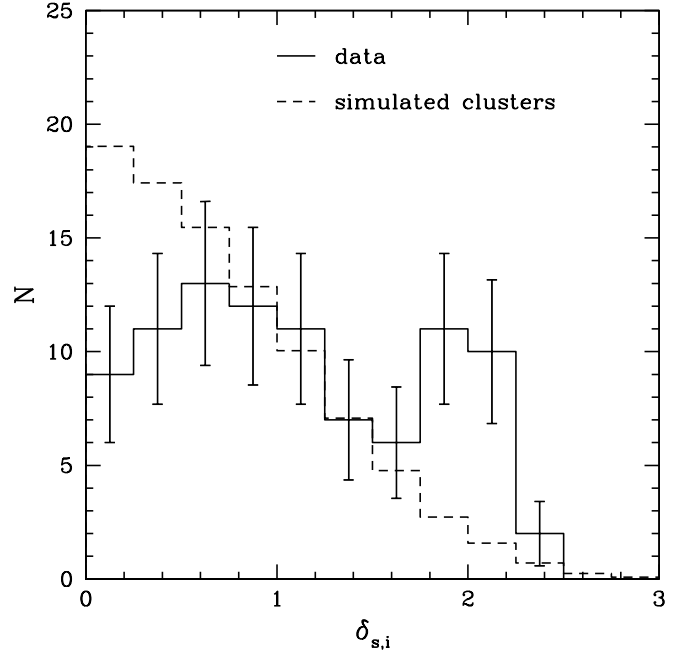}}
\caption
{Distribution of $\delta_{{\rm s},i}$ deviations of the
  Dressler-Schectman analysis for the 92 member galaxies. The solid
  line represents the observations with Poissonian errors. The dashed
  line is the distribution for the galaxies of simulated clusters,
  normalized to the observed number.}
\label{figdeltai10}
\end{figure}

To better understand the results of the DS-test described above, we
also considered two kinematical estimators alternative to the $\delta$
parameter of the DS-test, i.e. we considered separately the
contributes of the local mean $\delta_{\rm V}^2= [(N_{\rm
    nn}+1)/\sigma_{\rm V}^2](\overline {\rm V_l} - \overline {\rm
  V})^2]$, and dispersion $\delta_{\rm s}^2= [(N_{\rm
      nn}+1)/\sigma_{\rm V}^2](\sigma_{\rm V,l} - \sigma_{\rm V})^2]$
    (see, e.g.  Girardi et al. \cite{gir97}, Ferrari et
    al. \cite{fer03}).  When considering the $\mathrm{\delta_s}$
    estimator, we found evidence for a peculiar local velocity
    dispersion at the $99.2\%$ c.l..  Figure~\ref{figdssegno10} --
    lower panel -- shows the distribution on the sky of all galaxies,
    each marked by a circle: the larger the circle, the larger the
    deviation $\delta_{{\rm s},i}$ of the local velocity dispersion
    from the global cluster value.  Figure~\ref{figdssegno10} shows:
    i) as A1758N(NW) is well detected as a region of low $\sigma_{\rm
      V,l}$, ii) a few galaxies in the NE central region have high
    values of $\sigma_{\rm V,l}$. The first result is expected
    for/interpreted as a system having a relaxed core owing to
    circular orbits or galaxy merger phenomena (i.e., Girardi et
    al. \cite{gir96} and \cite{gir97}; Menci \& Fusco-Femiano
    \cite{men96}). The second result deserves more attention: we
    compared the distribution of individual $\delta_{{\rm s},i}$
    values of real galaxies with those of simulated clusters (Biviano
    et al. \cite{biv02}). Figure~\ref{figdeltai10} suggests that
    significantly high $\delta_{{\rm s},i}$ values for real data are
    those with $\delta_{{\rm s},i}>1.8$. A part from several galaxies
    with peculiarly low $\sigma_{\rm V,l}$, the 1.8 value only selects
    four galaxies with peculiarly high $\sigma_{\rm V,l}$ (the largest
    thick/red circles in Fig.~\ref{figdssegno10} -- lower panel,
    IDs~81, 90, 93, and 96), two of which are ELGs.

\begin{figure}
\centering 
\resizebox{\hsize}{!}{\includegraphics{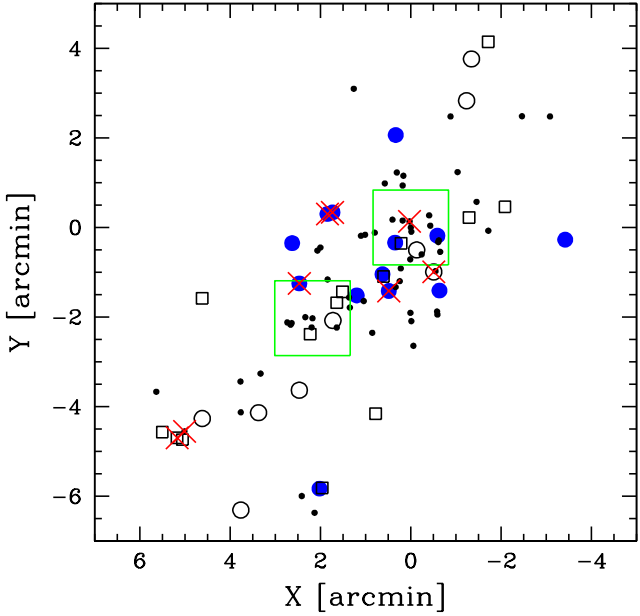}}
\caption
{Spatial distribution on the sky of the cluster galaxies, showing the
  four groups recovered by the weighted gap analysis. Solid/blue
  circles, open circles, open squares, and dots indicate the galaxies
  of GV1, GV2, GV3, and GV4, respectively.  The BCGN(NW) is taken as the
  cluster center.  Very large open green squares indicate the
  positions of BCGN(NW) and BCGN(SE). Large/red crosses indicate ELGs.}
\label{figxy1234}
\end{figure}

Then we searched for a possible physical meaning of the four
subclusters determined by the three weighted gaps. We compared the
spatial galaxy distributions of GV1, GV2, GV3, and GV4 two by two. GV1
galaxies are distributed in a round region between the NW and the SE
clumps, and do not follow the elongated shape of the sample.  We
verified that GV1 differs from GV3 (and from GV2+GV3) at the $94\%$
c.l. ($95\%$ c.l.)  according to the 2D Kolmogorov-Smirnov test
(2DKS-test, Fasano et al. \cite{fas87}, see Fig.~\ref{figxy1234}).

We finally resorted to the method devised by Serna \& Gerbal
(\cite{ser96}, hereafter Htree-method; see, e.g., Durret et
al. \cite{dur10} for a recent application). This method uses a
hierarchical clustering analysis to determine the relationship between
galaxies according to their relative binding energies.  The method
assumes a constant value for the mass-to-light ratio of galaxies and
Serna \& Gerbal (\cite{ser96}) suggested a value comparable to that of
clusters.  Here we took a value of $M/L_r$=150 \mll, as suggested by
large statistical studies (e.g., Girardi et al. \cite{gir00}; Popesso
et al. \cite{pop05}).

\begin{figure}
\centering 
\resizebox{\hsize}{!}{\includegraphics{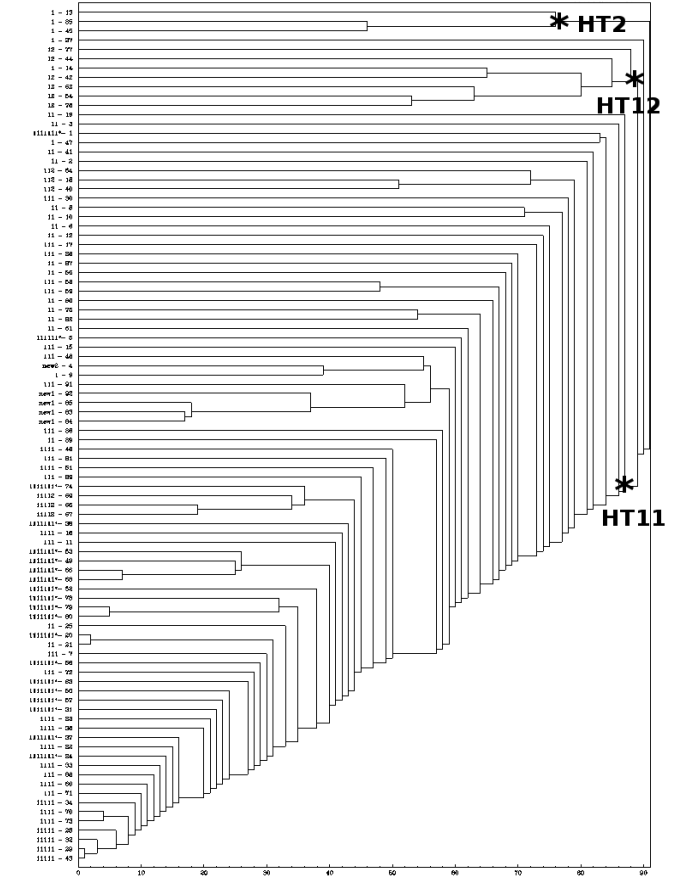}}
\caption
{Dendogram obtained through the Serna \& Gerbal (\cite{ser96})
  algorithm.  The abscissa is the binding energy (here in arbitrary
  units with the deepest negative energy levels on the left) while the
  catalog numbers of the various member galaxies are shown along the
  ordinate (IDm in Table~\ref{catalogA1758}).}
\label{a1758gerbal}
\end{figure}

\begin{figure}
\centering 
\resizebox{\hsize}{!}{\includegraphics{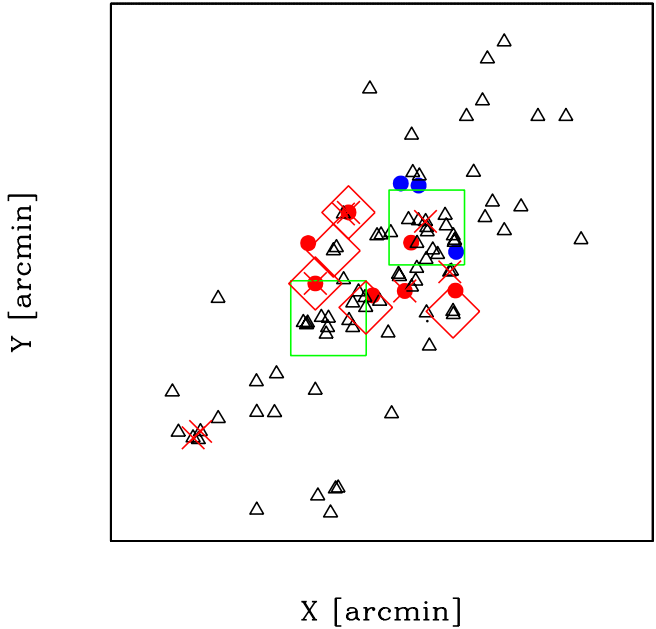}}
\caption
{Spatial distribution on the sky of the cluster galaxies showing the
  subsystems recovered by the Htree analysis. Solid blue and red
  circles indicate the galaxies of HT2 and HT12,
  respectively. Triangles indicate the galaxies of the main system
  (HT1).  The BCGN(NW) is taken as the cluster center.  Very
  large/green squares indicate the positions of BCGN(NW) and
  BCGN(SE). Large/red crosses indicate ELGs. Large red diamonds
  indicate dusty star-forming galaxies in the sample of Haines et
  al. (\cite{hai09}).  }
\label{figxyht}
\end{figure}

Figure~\ref{a1758gerbal} shows the resulting dendogram, where the
total energy appears horizontally.  Galaxy pairs and subgroups of
galaxies appear with a lower total energy. We note that there are no
important subsystems. In particular, the BCGN(SE) and BCGN(NW) lie at
the bottom of the potential well, although in two separate subsystems
(a pair and a quadruplet, respectively). Instead, there are several
small subsystems. We highlight here two that are situated at the
highest energy levels, which might largely bias the computation of the
global velocity dispersion. In Figure~\ref{a1758gerbal} we indicate
HT2, a triplet of close galaxies at high velocity ($\rm{v}>87\,000$
\kss), and HT12, which is formed by seven galaxies ($\rm{v}\sim
80\,000$ \ks) that lie in the central cluster region. Note that all
seven galaxies of HT12 are also members of the GV1 group detected in
our 1D analysis.  Other significant subsystems of HT11, here
interpreted as the main system, are small and generally formed by
close galaxies. We suspect that their detection might be very
sensitive to our incomplete spatial sampling and we do not discuss
them.  Figure~\ref{figxyht} shows the spatial distributions of HT2 and
HT12 with respect to the main system.

\section{Discussion and conclusions}
\label{discu}

The high value of the velocity dispersion $\sigma_{\rm
  V}=1329_{-116}^{+135}$ \ks is comparable to the values found for
hot, massive clusters (e.g., Mushotzky \& Scharf \cite{mus97}; Girardi
\& Mezzetti \cite{gir01}).  However, our estimate of $\sigma_{\rm V}$
is somewhat higher than the X-ray temperature of A1758N subclusters
under the assumption of the equipartition of energy density between
ICM and galaxies. Indeed, assuming that $kT_{\rm X}\sim 7$ keV, we
obtained $\beta_{\rm spec} \sim 1.24$ to be compared with $\beta_{\rm
  spec}=1$, where $\beta_{\rm spec}=\sigma_{\rm V}^2/(kT_{\rm X}/\mu
m_{\rm p})$ with $\mu=0.58$ the mean molecular weight and $m_{\rm p}$
the proton mass (see also Fig.~\ref{figprof}). Indeed, our
$\sigma_{\rm V}$ would instead predict a value of $kT_{\rm X}$=10.7
keV, which would be more in line with the temperature found in the
region beyond the two cores, where the gas is already shocked by the
merger and the X-ray temperature likely already enhanced to the value
of the whole forming cluster (David \& Kempner \cite{dav04} and
discussion therein). Below we discuss the bimodal nature of A1758N.

\subsection{Cluster structure and mass}
\label{stru}

In agreement with previous results (see Sect.~\ref{intro}), our 2D
analysis highlights two subclusters in the galaxy distribution about
3\arcmin ($\sim 0.75$ \hh) away from each other along the SE-NW
direction, hereafter A1758N(NW) and A1758N(SE).

\begin{figure}
\centering 
\resizebox{\hsize}{!}{\includegraphics{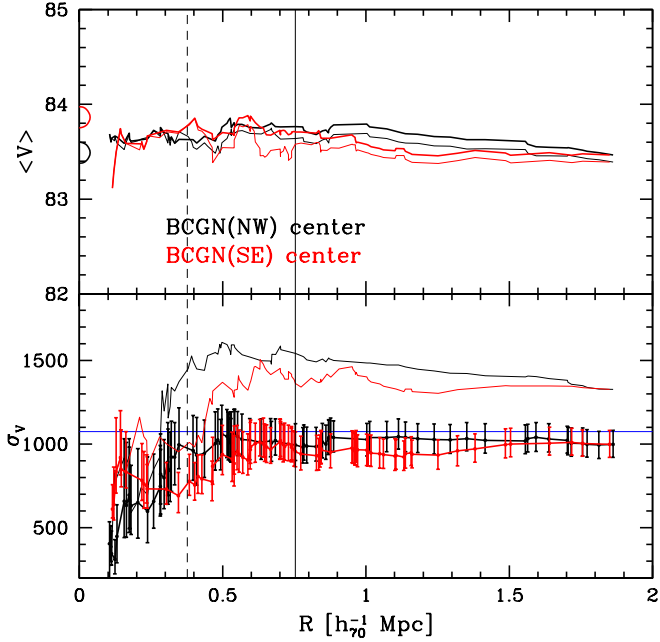}}
\caption
{ {\em Upper and lower panels:} integral profiles of mean velocity and
  LOS velocity dispersion, respectively. Black and red lines consider
  as centers BCGN(NW) and BCG(SE), respectively.  In both panels the
  vertical solid lines indicate the distance of the BCG of the close subcluster
  and the dashed lines indicate the half
  distance.  Thin and thick lines give the
  results for all galaxies and the main system HT11. In the upper panel,
  the two semi-circles give the velocity of the two BCGs. In the lower
  panel, the horizontal line represents the X-ray temperature 
as in Fig.~\ref{figprof}.  
}
\label{figprofNWSE}
\end{figure}

We cannot separate A1758N(NW) and A1758N(SE) from the velocity
information. Indeed, the small difference between the LOS velocities
of the two BCGs ($\siml 300$ \ks in the rest-frame) is by itself an
indication that the two subclusters have low LOS relative velocity. In
this study we were able to fully exploit our sample of 92 cluster
members. From i) the inspection of the galaxy velocity field in
Fig.~\ref{figdssegno10} (upper panel), ii) the absence of a velocity
gradient, iii) the absence of local velocity peculiarities via the
DS-test, we confirm that the velocity difference between the two
subclusters is very small.  This is clearly shown by
Fig.~\ref{figprofNWSE} -- upper panel, where the integral mean
velocities computed around A1758N(NW) and A1758N(SE) are very similar,
lower than the velocity difference between the two BCGs.

Taking into account the similar velocities of the two BCGs, the sharp
increase of the velocity dispersion profile within $0.5$ \h (see
Fig.~\ref{figprof} and Fig.~\ref{figprofNWSE} -- faint red line -- for
both the subclusters) is not directly explainable with the presence of
two systems at different mean velocity.  Rather, this might be due to
the presence of subclusters or individual galaxies infalling onto, or
escaping from, the cluster. Indeed, beyond the usual infall onto the
cluster from the large-scale structure, the recent merger of two
subclumps may result in outflying galaxies as shown in simulations
(e.g., Czoske et al. \cite{czo02}) and seen in a few clusters (see
e.g., the plume of outflying galaxies in Abell 3266 by Quintana et
al. \cite{qui96} and Flores et al. \cite{flo00}; the structure of
Cl0024+1654 by Czoske et al. \cite{czo02}). In particular, the
simulations reproducing Cl0024+1654 by Czoske et al. (\cite{czo02})
show that during a head-head collision at 3000 \kss, the outer
regions of the smaller cluster have become unbound and are streaming
radially away from the impact location with velocities perpendicular
to the encounter on the order of 1000 \kss.

The velocity information is useful for detecting the galaxy groups
connected with A1758N. Through our 3D Htree analysis we detected the
high velocity HT2 group and the low velocity HT12 group. The HT2 and
HT12 groups lie in the tails of the velocity distribution (which, in
fact, is a heavy tailed distribution; see Sect.~\ref{1d}). In
particular, HT12 is a subsample of GV1, the only group detected in the
1D analysis that also has a peculiar spatial distribution.  Spatially,
HT2 and HT11 lie in the central region roughly between the two BCGs,
where, in the NE, the DS analysis detects a region of high local
velocity dispersion. Figure~\ref{figprofNWSE} -- lower panel -- shows
the resulting profile of the velocity dispersion when only the main
system HT11 is considered: the effect is that the value of
$\sigma_{\rm V}$ decreases to $\sim 1000$ \kss.

As for individual/peculiar galaxies, we highlight the NAT galaxy
1330+507 (ID~105). Head-tail and NAT radio galaxies are most common in
clusters with perturbed X-ray morphologies and are probably produced
while radio galaxies move through cluster atmospheres at high
velocities (Burns et al. \cite{bur94}). Here the NAT galaxy was
rejected in the second step of our member selection procedure (the
galaxy represented by the square with the lowest velocity at
R$\sim$0.7 \h in Fig.~\ref{figprof} -- top panel) and therefore it is
likely to be a galaxy ``at the border'' of the phase-space
distribution of A1758N. The NAT, projected onto close BCGN(SE), points
toward A1758N(NW) (see Fig.~\ref{figimage}), which suggests that it is
infalling onto the cluster center.

In their analysis of galaxies with {\em Spitzer}/MIPS 24$\mu$m
emission in the region of A1758, Haines et al. (\cite{hai09}) stressed
that these dusty star-forming galaxies trace the cluster potential and
in particular the core of A1758N. This supports the hypothesis that
activity in clusters is triggered by the cluster-cluster merger
(i.e. is due to the passage of gas-rich galaxies through shocked
regions of the ICM or affected by time-dependent cluster tidal fields,
e.g., Roettiger et al. \cite{roe96}; Bekki \cite{bek99}; Bekki et
al. \cite{bek10}). However, they used photometric redshifts to
establish the A1758 membership, while we were able to use
spectroscopic redshifts to asses their result. We have redshifts for
twelve galaxies out of infrared luminous galaxies plotted in their
Fig.~1 (blue circles). Out of these twelve galaxies, five are cluster
members and they indeed lie in the core of A1758N, between the two
BCGs, thus supporting the conclusions of Haines et al. (\cite{hai09};
see large red diamonds in Fig.~\ref{figxyht}).  Moreover, another four
luminous infrared galaxies were rejected from the cluster members only
at the second step of our procedure, i.e. they lie ``at the border''
of the projected phase-space distribution shown in Fig.~\ref{figprof}
(top panel). Very interestingly, galaxies with emission lines have a
similar behavior: all 14 ELGs in our spectroscopic catalog are
connected to the cluster, eight of them belonging to the cluster and
six at the border (see Fig.~\ref{figprof}). In particular, six out of
the eight members are embedded in the cluster core (see
Fig.~\ref{figxyht}).  Thus, also the ELGs support the hypothesis that
activity in A1758N is triggered by the cluster-cluster merger. The
effect of the cluster merger on the galaxy population was also
suggested by Durret et al. (\cite{dur11}), who found an excess of very
bright galaxies and a bump at $M_{r^{\prime}}\sim -17$ in the observed
luminosity function with respect to a Schechter luminosity function.

Here we discuss the relative importance of the NW and SE subclusters.
From the galaxy clump luminosities, Okabe \& Umetsu (\cite{oka08})
suggested that A1758N(NW) is the primary component and A1758N(SE) is
the merging substructure, but they measured similar velocity
dispersions ($\sigma_{\rm V,SIS}\sim$ 600--700 \kss) from their
weak-lensing analysis. This conclusion also agrees with the fact that
A1758N(SE) shows an offset with the respective X-ray peak and
A1758N(NW) does not.  Our 2D-DEDICA analysis shows that A1758N(NW) is
richer than A1758N(SE), which suggests that A1758N(NW) is the main
structure. On the other hand, the SE peak shows a comparable density
(see Table ~\ref{tabdedica2d}). This supports the idea that A1758N(SE)
is the surviving, dense core of a merging structure.  The determination
of the velocity dispersion of subclusters is often a difficult
task. Figure~\ref{figprofNWSE} -- lower panel -- shows the
$\sigma_{\rm V}$ profiles using alternatively BCGN(NW) or BCGN(SE) as
centers.  As for A1758N(NW), the value of its $\sigma_{\rm V}$ is
progressively increasing -- which agrees with the small core with
peculiar low local $\sigma_{\rm V}$ shown by the DS-test -- out to
$\sim 1000$ \kss.  As for A1758N(SE), the value of its $\sigma_{\rm V}$ is
mostly stable at $\sim$ 700--800 \ks before increasing at about half
the distance from A1758N(NW).  Considering the large errors, there is no
significant difference between the values of $\sigma_{\rm V}$ of
A1758N(NW) and A1758N(SE).  However, in agreement with the idea that
A1758N(NW) is the main system, below we consider the nominal
values at the half distance, i.e. $\sigma_{\rm V,NW}\sim 1000$ \ks and
$\sigma_{\rm V,SE}\sim 800$ \ks for A1758N(NW) and A1758N(SE),
respectively.

Making the usual assumptions for the two subclusters (cluster
sphericity, dynamical equilibrium, coincidence of the galaxy and mass
distributions), we can compute virial global quantities.  We followed
the prescriptions of Girardi \& Mezzetti (\cite{gir01}, see also
Girardi et al. \cite{gir98}) and computed $R_{\rm vir}$ -- an estimate
for $R_{\rm 200}$ -- and the mass contained within this radius.  In
particular, we assume for the radius of the quasi-virialized region
$R_{\rm vir}=0.17\times \sigma_{\rm V}/H(z)$ \h (see Eq.~1 of Girardi
\& Mezzetti \cite{gir01} with the scaling of $H(z)$ of Eq.~ 8 of
Carlberg et al. \cite{car97} for $R_{200}$).  For the mass we used the
virial mass corrected for the surface pressure term correction
$M=M_{\rm V}-SPT=3\pi/2 \cdot \sigma_{\rm V}^2 R_{\rm PV}/G-SPT$
(Eq.~3 of Girardi \& Mezzetti \cite{gir01}), with the size estimate
$R_{\rm PV}$ computed using the full procedure based on an assumed
typical galaxy distribution and $SPT=0.2\times M_{\rm V}$.  In
practice, both $R_{\rm vir}$ and $M$ were computed on the basis of the
estimated velocity dispersion with the usual scaling-laws, where
$R_{\rm vir} \propto \sigma_{\rm V}$ and $M(<R_{\rm vir}) \propto
\sigma_{\rm V}^3$.  We computed $M_{\rm NW}(<R_{\rm vir}=2.1\hhh)=1.4$
\mqui and $M_{\rm SE}(<R_{\rm vir}=1.7\hhh)=0.7$ \mqui for the NW and
SE subsystems, respectively.  We computed a mass $M_{\rm
  sys}(<R=2.1\hhh)\simg 2.1$ \mqui for the whole system, depending on
how far we extrapolated the mass of A1758N(SE) outside its $R_{\rm
  vir}$. This value agrees with the value we obtained assuming the
whole system to be relaxed $M(<R_{\rm vir}=2.8\hhh)= 3\pm1$ \mqui when
rescaled to the smaller $R=2.1$ \h radius.

As for the cluster structure, finally we discuss our result of faint
galaxies tracing a peak intermediate between the two BCGs, somewhat
close to the secondary X-ray peak, while luminous galaxies trace two
subclusters (see Fig.~\ref{figk2cfht}).  This phenomenon of luminosity
segregation can be connected with the cluster internal
dynamics. Indeed, very appealingly, galaxies of different luminosity
could trace the dynamics of cluster mergers in a different way.  A
first evidence was provided by Biviano et al.  (\cite{biv96}): they
found that the two central dominant galaxies of the Coma cluster are
surrounded by luminous galaxies, accompanied by the two main X-ray
peaks, while the distribution of faint galaxies tends to form a
structure not centered with one of the two dominant galaxies, but is
instead coincident with a secondary peak detected in X-ray data.  The
observational scenario of Abell 209 has some peculiar luminosity
segregation, too (Mercurio et al. \cite{mer03}).  Therefore, following
Biviano et al. (\cite{biv96}), we speculate that the merging is in a
post-merging phase, where faint galaxies trace the forming structure
of the cluster in formation, while more luminous galaxies still trace
the remnant of the core-halo structure of pre-merging subclusters.

\subsection{Merger kynematics and diffuse radio sources}
\label{rad}

Although member galaxies can give some pieces of evidence in favor of
a post-merger phase, the stringent proof is given by the X-ray
analysis of David \& Kempner (\cite{dav04}, {\em Chandra} and {\em
  XMM-Newton} data) who detected shock-heated gas around the cores
of the two subclusters.  In this study we obtained some important
properties of the merger: the LOS relative velocity is $\siml 300$ \ks
and the merger is a major merger with a mass ratio of $\sim$
(2:1). Moreover, in agreement with the results of Pinkney et
al. (\cite{pin96}), the fact that we clearly detected the subclusters in
2D, but not in the 1D and 3D analyses, suggests that the plane of the
cluster merger is mostly perpendicular to the LOS. This would also
explain the good observability of the X-ray morphological features.

Figure~\ref{figimage} shows the comparison between the positions of
the density peaks in the galaxy distribution obtained in this work and
the results from other wavelengths. In agreement with the fact that
elongated radio halos normally follow the merging direction (e.g., the
``bullet'' cluster by Markevitch et al. \cite{mar02}; Abell 520 by
Girardi et al. \cite{gir08}; Abell 754 by Macario et
al. \cite{mac11}), the halo of A1758N is clearly elongated along the
NW-SE direction. As for the two relics, the eastern one (R1 in
Fig.~\ref{figimage}) is not perpendicular to the merging axis, as it
is typically found in double relic clusters (e.g., Abell 3667
Roettiger et al. \cite{roe99}; Kempner \& Sarazin \cite{kem01}). This
supports the idea that A1758N is not a head-on merger (see below, but
see Bonafede et al. \cite{bon09} and Boschin et al. \cite{bos10} for
alternative explanations of non-symmetric radio relics in Abell 2345).
More quantitatively, we can use our results to examine A1758N with
respect to the observed scaling relation between $P_{\rm 1.4GHz}$, the
halo radio power, and the total cluster mass (as determined within
3\hh, Govoni et al. \cite{gov01}).  Considering the value of $P_{\rm
  1.4GHz}=9.3\,10^{23}$ W Hz$^{-1}$ by Giovannini et
al. (\cite{gio09}) and our $M_{\rm sys}=2$--3 \mquii, A1758N results
to be an under-radio-luminous cluster (see Fig.~18 of Govoni et
al. \cite{gov01}).  However, it should be said that A1758N is also
very peculiar because of the presence of the two close relics. When
considering the total diffuse emission radio+relics $P_{\rm
  1.4GHz}=4\,10^{24}$ W Hz$^{-1}$ (Giovannini et al. \cite{gio09}),
there is no longer a discrepancy.

The presence of the diffuse radio emissions suggests that A1758N
probably is a recent merger because they are expected to have a short
life time, i.e. on the order of a few fractions of Gyr (e.g.,
Giovannini \& Feretti \cite{gio02}; Skillman et al. \cite{ski11}). For
radio halos, present estimates of the time elapsed after the core-core
crossing are very short (e.g., 0.1-0.2 Gyr in the ``bullet'' cluster
by Markevitch et al. \cite{mar02}; 0.2-0.3 in Abell 520 by Girardi et
al. \cite{gir08}). Times are somewhat longer for the radio relics (1
Gyr in Abell 3667 by Roettiger et al. \cite{roe99}), but they refer to
clusters whose relics are separated by $\simg$ 2 \hh, therefore we
expect that the times for A1758N relics are shorter.  An independent
evidence of the early age of the cluster merger comes from the
presence of the bump of highly star-forming galaxies in the cluster
core, being the duration of the strong starburst induced by the
cluster merger of $<0.5$ Gyr (Bekki \cite{bek99}).

\begin{figure}
\centering
\resizebox{\hsize}{!}{\includegraphics{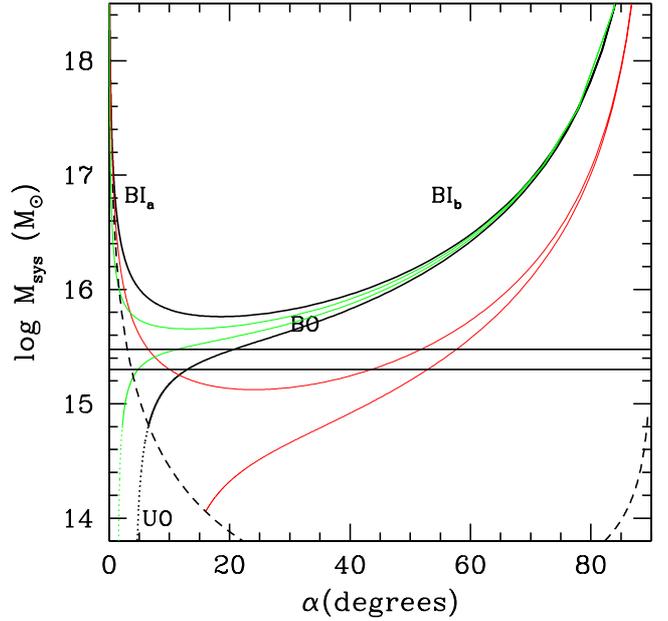}}
\caption
{System mass vs. projection angle for bound and unbound solutions
  (thick solid and thick dashed curves, respectively) of the two-body
  model applied to the A1758N(NW) A1758N(SE) subclusters.  Thick/black
  lines refer to $t=0.2$ Gyr and $V_{\rm rf,LOS}=300$ \kss.  Labels
  BI$_{\rm a}$ and BI$_{\rm b}$ indicate the bound and incoming, i.e.,
  collapsing solutions (solid curve). Labels BO and UO indicate the
  bound outgoing, i.e., expanding solutions and unbound outgoing
  solutions (solid curve going on in the dotted curve,
  respectively). The horizontal lines give the range of observational
  values of the mass system.  The dashed curve separates bound
  and unbound regions according to the Newtonian criterion (above and
  below the thin dashed curve, respectively).  The red/faint lines
  refer to $t=0.5$ Gyr and $V_{\rm rf,LOS}=300$ \kss.  The green/faint
  lines refer to $t=0.2$ Gyr and $V_{\rm rf,LOS}=100$ \kss.  }
\label{figbim}
\end{figure}

On the basis of our results we investigated the relative dynamics of
A1758N and A1758S using different analytic approaches, that are based
on an energy integral formalism in the framework of locally flat
spacetime and Newtonian gravity (e.g., Beers et al. \cite{bee82}).
The values of the relevant observable quantities for the two-clumps
system are the relative LOS velocity in the rest frame, $V_{\rm
  r}=300$ \ks (which is an upper limit); the projected linear distance
between the two clumps, $D\sim0.8$ \hh; the mass of the system $M_{\rm
  sys}=2$--3 \mquii.  First, we considered the Newtonian criterion for
gravitational binding stated in terms of the observables as $V_{\rm
  r}^2D\leq2GM_{\rm sys}\sin^2\alpha \cos\alpha$, where $\alpha$ is
the projection angle between the plane of the sky and the line
connecting the centers of two clumps.  The dashed curve in
Fig.~\ref{figbim} separates the bound and unbound regions according to
the Newtonian criterion (above and below the curve, respectively). The
system is bound between $\sim 3\degree$ and $90\degree$; the
corresponding probability, computed considering the solid angles
(i.e., $\int^{90}_{5} \cos\alpha\,d\alpha$), is $95\%$.  We also
considered the implemented criterion $V_{\rm
  r}^2D\leq2GM\sin^2\alpha_{\rm V} \cos\alpha_{\rm D}$, which
introduces different angles for projection of distance and velocity,
not assuming a strictly radial motion between the clumps (Hughes et
al. \cite{hug95}).  We obtained a binding probability of $94\%$.

In the simple case of a strictly radial motion we can check if the
merger kinematics agrees with the idea of a very recent merger because
the simple linear-orbit two-body model connects the time elapsed from
the the core-core passage $t$ and the projection angle $\alpha$.
Figure~\ref{figbim} shows the bimodal-model solutions as a function of
$\alpha$. The thick/black lines show the case with $V_{\rm
  rf,LOS}=300$ \kss and the time from the core-core passage $t=0.2$
Gyr. There is a bound outgoing solution (BO) with $\alpha \sim$
15--20$\degree$ that leads to a deprojected relative velocity of
$V_{\rm rf}\sim$ 1150--900 \kss.  The effect of decreasing the
relative velocity is that of decreasing the projection angle, e.g.
$V_{\rm rf,LOS}=100$ \ks leads to $\alpha \sim$ 5--10$\degree$ and
thus to $V_{\rm rf}\simg$ 1150--600 \kss.  The effect of increasing
the age of the merger is that of increasing the projection angle and
decreasing the relative velocity, e.g.  $t=0.5$ Gyr leads to $\alpha
\sim$ 55--60$\degree$ and thus to $V_{\rm rf}\sim$ 350 \ks (here there
are also two symmetric bound incoming solutions). Therefore only a
very recent merger agrees with the evidence that the merger plane is
mostly perpendicular to the LOS.  Moreover, we have to consider the
limits on the relative velocity indirectly recovered from X-ray data.
In order to produce the shock waves, which e.g. originate the relics,
we need a Mach number ${\cal M}>1$ and David \& Kempner (\cite{dav04})
computed ${\cal M}\siml 1.15$ from X-ray data. The Mach number is
defined to be ${\cal M}={\rm v}_{\rm s}/c_{\rm s}$, where ${\rm
  v}_{\rm s}$ is the velocity of the shock and $c_{\rm s}$ in the
sound speed in the pre-shock gas (see e.g., Sarazin \cite{sar02} for a
review). Assuming for the sound speed in the pre-shock gas the value
obtained from the X-ray temperature of David \& Kempner (6-9 keV),
i.e. ${\rm v}_{\rm s}=1000$--1200 \kss, and ${\rm v}_{\rm s} \sim
V_{\rm rf}$, we expect $V_{\rm rf}$ to be in the range of 1000-1300
\kss, which agrees with models with a very recent merger.

The obvious limit of the model described above is that the merger is
treated as a head-on merger. Although general arguments suggest
small impact parameters for cluster mergers ($\sim$0.2 \hh, see
Sarazin \cite{sar02}), it seems that this is not the case for A1758N.
From the morphological details of the X-ray surface brightness image,
David \& Kempner (\cite{dav04}) deduced that the NW subcluster is
moving north and the SE subcluster is moving southeast and that
A1758N is in the late stage of a large impact parameter merger: their
Fig.~9 shows the proposed orbits in the merging plane.  However, these
orbits are not supported by the two elongated regions of high
metallicity likely left by ram-pressure stripping during the merger
(Sect.~4 and Fig.~13 of Durret et al. \cite{dur11}): the NW feature is
elongated along NW-SE as in the case of a head-on merger, while the
SW feature has a clear tail perpendicular to the NW-SE direction as
in the case of an off-axis merger. Thus, to date, we are far from
knowing the details of the possible orbits and from investigating a
more complex kinematics.

Our general conclusions about A1758 agree with other typical clusters
that show extended diffuse radio emissions, i.e. we are dealing with a
massive cluster with a major ongoing merger (like, e.g., Abell 2744;
see Boschin et al. \cite{bos06}) and, indeed, less massive merging
systems do not seem to host extended radio emissions (e.g. Abell 2146
by Russell et al. \cite{rus11}; see also Rodr\'iguez-Gonz\'alvez et
al. \cite{rod11}).

\begin{acknowledgements}

WB dedicates this study to the memory of his father, Andriano Boschin,
deceased on November 15, 2011. We acknowledge Federica Govoni and
Gabriele Giovannini for the VLA radio image they kindly provided us
and their useful comments. We also thank the anonymous referee for
his/her stimulating comments and suggestions.

This publication is based on observations made on the island of La
Palma with the Italian Telescopio Nazionale Galileo (TNG). The TNG is
operated by the Fundaci\'on Galileo Galilei -- INAF (Istituto
Nazionale di Astrofisica) and is located in the Spanish Observatorio
of the Roque de Los Muchachos of the Instituto de Astrofisica de
Canarias.

This research has made use of the NASA/IPAC Extragalactic Database
(NED), which is operated by the Jet Propulsion Laboratory, California
Institute of Technology, under contract with the National Aeronautics
and Space Administration.

This research has made use of archival data obtained at the
Canada-France-Hawaii Telescope (CFHT), which is operated by the
National Research Council of Canada, the Institut National des
Sciences de l'Univers of the Centre National de la Recherche
Scientifique of France, and the University of Hawaii.

This research has made use of the galaxy catalog of the Sloan Digital
Sky Survey (SDSS). Funding for the SDSS has been provided by the
Alfred P. Sloan Foundation, the Participating Institutions, the
National Aeronautics and Space Administration, the National Science
Foundation, the U.S. Department of Energy, the Japanese
Monbukagakusho, and the Max Planck Society. The SDSS Web site is
http://www.sdss.org/.

The SDSS is managed by the Astrophysical Research Consortium for the
Participating Institutions. The Participating Institutions are the
American Museum of Natural History, Astrophysical Institute Potsdam,
University of Basel, University of Cambridge, Case Western Reserve
University, University of Chicago, Drexel University, Fermilab, the
Institute for Advanced Study, the Japan Participation Group, Johns
Hopkins University, the Joint Institute for Nuclear Astrophysics, the
Kavli Institute for Particle Astrophysics and Cosmology, the Korean
Scientist Group, the Chinese Academy of Sciences (LAMOST), Los Alamos
National Laboratory, the Max-Planck-Institute for Astronomy (MPIA),
the Max-Planck-Institute for Astrophysics (MPA), New Mexico State
University, Ohio State University, University of Pittsburgh,
University of Portsmouth, Princeton University, the United States
Naval Observatory, and the University of Washington.

\end{acknowledgements}


\begin{thebibliography}{}

\bibitem[1989]{abe89} Abell, G. O., Corwin, H. G. Jr., \& Olowin, R. P. 1989, \apjs, 70, 1

\bibitem[1994]{ash94} Ashman, K. M., Bird, C. M., \& Zepf, S. E. 1994, \aj, 108, 2348

\bibitem[1994]{bar94} Bardelli, S., Zucca, E., Vettolani, G., et al. 1994, \mnras, 267, 665 

\bibitem[2009]{bar09} Barrena, R., Girardi, M., Boschin, W., \& Das\'i, M. 2009, \aap, 503, 357

\bibitem[1990]{bee90} Beers, T. C., Flynn, K., \& Gebhardt, K. 1990, \aj, 100, 32

\bibitem[1991]{bee91} Beers, T. C., Forman, W., Huchra, J. P., Jones, C., \& Gebhardt, K. 1991, \aj, 102, 1581

\bibitem[1992]{bee92} Beers, T. C., Gebhardt, K., Huchra, J. P., et al. 1992, \apj, 400, 410

\bibitem[1982]{bee82} Beers, T. C., Geller, M. J., \& Huchra, J. P. 1982, \apj, 257, 23

\bibitem[1999]{bek99} Bekki, K. 1999, \apj, 510, L15

\bibitem[2010]{bek10} Bekki, K., Owers, M. S., \& Couch, W. J. 2010, \apjl, 718, L27

\bibitem[1993]{bir93} Bird, C. M., \& Beers, T. C., 1993, \aj, 105, 1596

\bibitem[1996]{biv96} Biviano, A., Durret, F., Gerbal, D., et al. 1996, \aap, 311, 95

\bibitem[2002]{biv02} Biviano, A., Katgert, P., Thomas, T., \& Adami, C. 2002, \aap, 387, 8

\bibitem[2009]{bon09} Bonafede, A., Feretti, L., Giovannini, G., et al. 2009, \aap, 503, 707

\bibitem[2010]{bos10} Boschin, W., Barrena, R., \& Girardi, M. 2010, \aap, 521, A78

\bibitem[2006]{bos06} Boschin, W., Girardi, M., Spolaor, M., \& Barrena, R. 2006, \aap, 449, 461

\bibitem[2009]{bru09} Brunetti, G., Cassano, R., Dolag, K., \& Setti, G. 2009, \aap, 507, 661

\bibitem[2002]{buo02} Buote, D. A. 2002, in ``Merging Processes in Galaxy Clusters'', eds. L. Feretti, I. M. Gioia, \& G. Giovannini (The Netherlands, Kluwer Ac. Pub.): Optical Analysis of Cluster Mergers

\bibitem[1994]{bur94} Burns, J. O., Roettiger, K., Ledlow, M., \& Klypin, A. 1994, \apj, 427, L87

\bibitem[1997]{car97} Carlberg, R. G., Yee, H. K. C., \& Ellingson, E. 1997, \apj, 478, 462

\bibitem[2005]{cas05} Cassano, R., \& Brunetti, G. 2005, \mnras, 357, 1313

\bibitem[2010a]{cas10a} Cassano, R., Brunetti, G., R\"ottgering, H. J. A., \& Br\"uggen, M. 2010a, \aap, 509, A68

\bibitem[2006]{cas06} Cassano, R., Brunetti, G., \& Setti, G. 2006, \mnras, 369, 1577

\bibitem[2010b]{cas10b} Cassano, R., Ettori, S., Giacintucci, S., et al. 2010b, \apjl, 721, L82

\bibitem[2002]{czo02} Czoske, O., Moore, B., Kneib, J.-P., \& Soucail, G. 2002, \aap, 386, 31

\bibitem[2002]{dah02} Dahle, H., Kaiser, N., Irgens, R. J., Lilje, P. B., \& Maddox, S. J. 2002, \apjs, 139, 313

\bibitem[1980]{dan80} Danese, L., De Zotti, C., \& di Tullio, G. 1980, \aap, 82, 322

\bibitem[2004]{dav04} David, L. P., \& Kempner, J. 2004, \apj, 613, 831

\bibitem[1996]{den96} den Hartog, R., \& Katgert, P. 1996, \mnras, 279, 349

\bibitem[1988]{dre88} Dressler, A., \& Shectman, S. A. 1988, \aj, 95, 985

\bibitem[2010]{dur10} Durret, F., Lagan\'a, T. F., Adami, C. \& Bertin, E. 2010, \aap, 517, A94

\bibitem[2011]{dur11} Durret, F., Lagan\'a, T. F., \& Haider, M. 2011, \aap, 529, A38

\bibitem[1998]{ebe98} Ebeling, H., Edge, A. C., B\"ohringer, H., et al. 1998, \mnras, 301, 881

\bibitem[1994]{ell94} Ellingson, E., \& Yee, H. K. C. 1994, \apjs, 92, 33

\bibitem[1998]{ens98} Ensslin, T. A., Biermann, P. L., Klein, U., \& Kohle, S. 1998, \aap, 332, 395

\bibitem[2001]{ens01} Ensslin, T. A., \& Gopal-Krishna 2001, \aap, 366, 26

\bibitem[2011]{ens11} Ensslin, T. A., Pfrommer, C., Miniati, F., \& Subramanian, K. 2011, \aap, 527, A99

\bibitem[1996]{fad96} Fadda, D., Girardi, M., Giuricin, G., Mardirossian, F., \& Mezzetti, M. 1996, \apj, 473, 670

\bibitem[1987]{fas87} Fasano, G., \& Franceschini, A. 1987, \mnras, 225, 155

\bibitem[1999]{fer99} Feretti, L. 1999, MPE Report No. 271

\bibitem[2002a]{fer02a} Feretti, L. 2002a, The Universe at Low Radio
  Frequencies, Proceedings of IAU Symposium 199, held 30 Nov -- 4 Dec
  1999, Pune, India. Edited by A. Pramesh Rao, G. Swarup, and
  Gopal-Krishna, 2002., p.133

\bibitem[2005]{fer05a} Feretti, L. 2005, X-Ray and Radio Connections
(eds. L. O. Sjouwerman and K. K. Dyer). Published electronically by
NRAO, http://www.aoc.nrao.edu/events/xraydio. Held 3-6 February 2004
in Santa Fe, New Mexico, USA

\bibitem[2005]{fer05b} Feretti, L., Schuecker, P., B\"ohringer, H., Govoni, F., \& Giovannini, G. 2005, \aap, 444, 157

\bibitem[2002b]{fer02b} Feretti, L., Gioia I. M., and Giovannini
G. eds., 2002b, Astrophysics and Space Science Library, vol. 272, 
``Merging Processes in Galaxy Clusters'', Kluwer Academic Publisher,
The Netherlands

\bibitem[2003]{fer03} Ferrari, C., Maurogordato, S., Cappi, A., \& Benoist, C. 2003, \aap, 399, 813

\bibitem[2000]{flo00} Flores, R. A., Quintana, H., \& Way, M. J. 2000, \apj, 532, 206

\bibitem[2009]{gio09} Giovannini, G., Bonafede, A., Feretti, L., et al.  2009, \aap, 507, 1257

\bibitem[2002]{gio02} Giovannini, G., \& Feretti, L. 2002, in
``Merging Processes in Galaxy Clusters'', eds. L. Feretti,
I. M. Gioia, \& G. Giovannini (The Netherlands, Kluwer Ac. Pub.):
Diffuse Radio Sources and Cluster Mergers

\bibitem[2006]{gio06} Giovannini, G., Feretti, L., Govoni, F., et
  al. 2006, Astron. Nachr., 327, 563

\bibitem[1999]{gio99} Giovannini, G., Tordi, M., \& Feretti, L. 1999, New Astronomy, 4, 141

\bibitem[2007]{gir07} Girardi, M., Barrena, R., \& Boschin, W. 2007,
  Contribution to ``Tracing Cosmic Evolution with Clusters of
  Galaxies: Six Years Later'' conference --
  http://adlibitum.oat.ts.astro.it/girardi/darc/sestomgirardi.pdf

\bibitem[2008]{gir08} Girardi, M., Barrena, R., Boschin, W., \& Ellingson, E. 2008, \aap, 491, 379

\bibitem[2002]{gir02} Girardi, M., \& Biviano, A. 2002, in ``Merging
Processes in Galaxy Clusters'', eds. L. Feretti, I. M. Gioia, \&
G. Giovannini (The Netherlands, Kluwer Ac. Pub.): Optical Analysis of
Cluster Mergers

\bibitem[2000]{gir00} Girardi, M., Borgani, S., Giuricin, G., Mardirossian, F., \& Mezzetti, M. 2000, \apj, 530, 62

\bibitem[1997]{gir97} Girardi, M., Escalera, E., Fadda, D., et al. 1997, \apj, 482, 11

\bibitem[1996]{gir96} Girardi, M., Fadda, D., Giuricin, G. et al. 1996, \apj, 457, 61

\bibitem[1998]{gir98} Girardi, M., Giuricin, G., Mardirossian, F., Mezzetti, M., \& Boschin, W. 1998, \apj, 505, 74

\bibitem[2001]{gir01} Girardi, M., \& Mezzetti, M. 2001, \apj, 548, 79

\bibitem[2002]{got02} Goto, T., Sekiguchi, M., Nichol, R. C., et al. 2002, \aj, 123, 1807

\bibitem[2001]{gov01} Govoni, F., Feretti, L., Giovannini, G., et al. 2001, \aap, 376, 803

\bibitem[2009]{gwy09} Gwyn, S. D. 2009, in Astronomical Data Analysis Software and Systems XVIII, ASP Conf. Ser., 411, 123 

\bibitem[2009]{hai09} Haines, C. P., Smith, G. P., Egami, E., et al. 2009, \mnras, 396, 1297

\bibitem[2009]{har09} Hart, Q. N., Stocke, J. T., \& Hallman, E. J. 2009, \apj, 705, 854

\bibitem[2004]{hoe04} Hoeft, M., Br\"uggen, M., \& Yepes, G. 2004, \mnras, 347, 389

\bibitem[1995]{hug95} Hughes, J. P., Birkinshaw, M., \& Huchra, J. P. 1995, \apjl, 448, 93

\bibitem[2004]{kem04} Kempner, J. C., Blanton, E. L., Clarke, T. E.
  et al. 2004, Proceedings of the conference ``The Riddle of Cooling
  Flows in Galaxies and Clusters of Galaxies'', eds. T. H. Reiprich,
  J. C. Kempner, \& N. Soker, e-print arXiv astro-ph/0310263

\bibitem[2001]{kem01} Kempner, J. C., \& Sarazin, C. L. 2001, \apj, 548, 639 

\bibitem[1992]{ken92} Kennicutt, R. C. 1992, \apjs, 79, 225

\bibitem[2010]{kes10} Keshet, U., \& Loeb, A. 2010, \apj, 722, 737

\bibitem[2007]{lop07} Lopes, P. A. A. 2007, \mnras, 380, 1608

\bibitem[2009]{lu09} Lu, T., Gilbank, D. G., Balogh, M. L., \& Bognat, A. 2009, \mnras, 399, 1858

\bibitem[2000]{lub00} Lubin, L. M., Brunner, R., Metzger, M. R., Postman, M., \& Oke, J. B. 2000, \apjl, 531, 5

\bibitem[1992]{mal92} Malumuth, E. M., Kriss, G. A., Dixon, W. Van Dyke, Ferguson, H. C., \& Ritchie, C. 1992, \aj, 104, 495

\bibitem[2002]{mar02} Markevitch, M., Gonzalez, A. H., David, L., et al. 2002, \apjl, 567, L27

\bibitem[2011]{mac11} Macario, G., Markevitch, M., Giacintucci, S., et al. 2011, \apj, 728, 82

\bibitem[2003]{mer03} Mercurio, A., Girardi, M., Boschin, W., Merluzzi, P., \& Busarello, G. 2003, \aap, 397, 431

\bibitem[1996]{men96} Menci, N., \& Fusco-Femiano, R. 1996, \apj, 472, 46

\bibitem[1997]{mus97} Mushotzky, R. F., \& Scharf, C. A. 1997, \apj, 482, L13

\bibitem[1985]{ode85} O'Dea, C. P., \& Owen, F. N. 1985, \aj, 90, 927

\bibitem[2008]{oka08} Okabe, N., \& Umetsu, K. 2008, \pasj, 60, 345

\bibitem[1996]{pin96} Pinkney, J., Roettiger, K., Burns, J. O., \& Bird, C. M. 1996, \apjs, 104, 1

\bibitem[1993]{pis93} Pisani, A. 1993, \mnras, 265, 706

\bibitem[1996]{pis96} Pisani, A. 1996, \mnras, 278, 697

\bibitem[2005]{pop05} Popesso, P., Biviano, A., B\"ohringer, H., Romaniello, M., \&  Voges, W. 2005, \aap, 433, 431

\bibitem[2000]{qui00} Quintana, H., Carrasco, E. R., \& Reisenegger, A. 2000, \aj, 120, 511

\bibitem[1999]{qui96} Quintana, H., Ram\'irez, A., \& Way, M. J. 1996, \aj, 112, 360

\bibitem[2012]{rag12} Ragozzine, B., Clowe, D., Markevitch, M., Gonzalez, A. H., \& Brada$\check{\rm c}$, M. 2012, \apj, 744, 94

\bibitem[1998]{riz98} Rizza, E., Burns, J. O., Ledlow, M. J., et al. 1998, \mnras, 301, 328

\bibitem[2003]{riz03} Rizza, E., Morrison, G. E., Owen, F. N., et al. 2003, \aj, 126, 119

\bibitem[2011]{rod11} Rodr\'iguez-Gonz\'alvez, C., Olamaie, M., Davies, M. L., et al. 2011, \mnras, 414, 3751

\bibitem[1996]{roe96} Roettiger, K., Burns, J. O., \& Loken, C. 1996, \apj, 473, 651

\bibitem[1999]{roe99} Roettiger, K., Burns, J. O., \& Stone, J. M. 1999, \apj, 518, 603

\bibitem[1997]{roe97} Roettiger, K., Loken, C., \& Burns, J. O. 1997, \apjs, 109, 307

\bibitem[2011]{rus11} Russell, H. R., van Weeren, R. J., Edge, A. C., et al. 2011, \mnras, 417, L1

\bibitem[2002]{sar02} Sarazin, C. L. 2002, in ``Merging Processes in
Galaxy Clusters'', eds. L. Feretti, I. M. Gioia, \& G. Giovannini (The
Netherlands, Kluwer Ac. Pub.): The Physics of Cluster Mergers

\bibitem[2001]{sch01} Schuecker, P., B\"ohringer, H., Reiprich, T. H., \& Feretti, L. 2001, \aap, 378, 408

\bibitem[1996]{ser96} Serna, A., \& Gerbal, D. 1996, \aap, 309, 65

\bibitem[2011]{ski11} Skillman, S. W., Hallman, E. J., O'Shea, B. W., et al. 2011, \apj, 735, 96 

\bibitem[1987]{str87} Struble, M. F. \& Rood, H. J. 1987, \apjs, 63, 555

\bibitem[1982]{tho82} Thompson, L. A. 1982, in IAU Symposium 104,
  ``Early Evolution of the Universe and the Present Structure'',
  eds. G.O. Abell and G. Chincarini (Dordrecht: Reidel)

\bibitem[1979]{ton79} Tonry, J., \& Davis, M. 1979, \apj, 84, 1511

\bibitem[1993]{tri93} Tribble, P. C. 1993, \mnras, 261, 57

\bibitem[2008]{ven08} Venturi, T., Giacintucci, S., Macario, G., et al. 2008, \aap, 484, 327

\bibitem[1978]{wai78} Wainer, H., \&  Schacht, S. 1978, Psychometrika, 43, 203

\bibitem[1990]{wes90} West, M. J., \& Bothun, G. D. 1990, \apj, 350, 36

\end{thebibliography}
\end{document}